\documentclass[11pt,a4paper]{article}
\pdfoutput=1
\usepackage{jinstpub}
\usepackage[utf8]{inputenc}
\usepackage[T1]{fontenc}
\usepackage{mathtools}
\usepackage{subfigure}
\usepackage{siunitx}
\usepackage{xfrac}
\usepackage{physics}
\usepackage[noabbrev]{cleveref}
\usepackage{lineno}

\title{MAGIS-100 Environmental Characterization and Noise Analysis}

\author[a,b,*]{J.~Mitchell,%
\note[*]{Corresponding author.}}
\author[c]{T.~Kovachy,}
\author[d]{S.~Hahn,}
\author[d]{P.~Adamson}
\author[b,d]{and S.~Chattopadhyay}

\affiliation[a]{Cavendish Laboratory, University of Cambridge, \\Cambridge, U.K.}
\affiliation[b]{Department of Physics, Northern Illinois University, \\DeKalb, Illinois 60115, U.S.A.}
\affiliation[c]{Center for Fundamental Physics, Northwestern University, \\Evanston, Illinois 60208, U.S.A.}
\affiliation[d]{Fermi National Accelerator Laboratory, \\Batavia, Illinois 60510, U.S.A.}

\emailAdd{jm2427@cam.ac.uk}

\abstract{We investigate and analyze site specific systematics for the MAGIS-100 atomic interferometry experiment at Fermi National Accelerator Laboratory. As atom interferometers move out of the laboratory environment passive and active mitigation for noise sources must be implemented. To inform the research and development of the experiment design, we measure ambient temperature, humidity, and vibrations of the installation site. We find that temperature fluctuations will necessitate enclosures for critical subsystems and a temperature controlled laser room for the laser system. We also measure and analyze the vibration spectrum above and below ground for the installation site. The seismic vibration effect of gravity gradient noise is also modeled using input from a low-noise seismometer at multiple locations and a mitigation scheme is studied using a stochastic simulation and characterized by a suppression factor.}

\date{\today}

\begin{document}
\maketitle

\flushbottom

\section{Background}
\label{sec:background}

\begin{figure}
  \centering
  \includegraphics[trim={0 1cm 0 0},clip,width=0.55\textwidth]{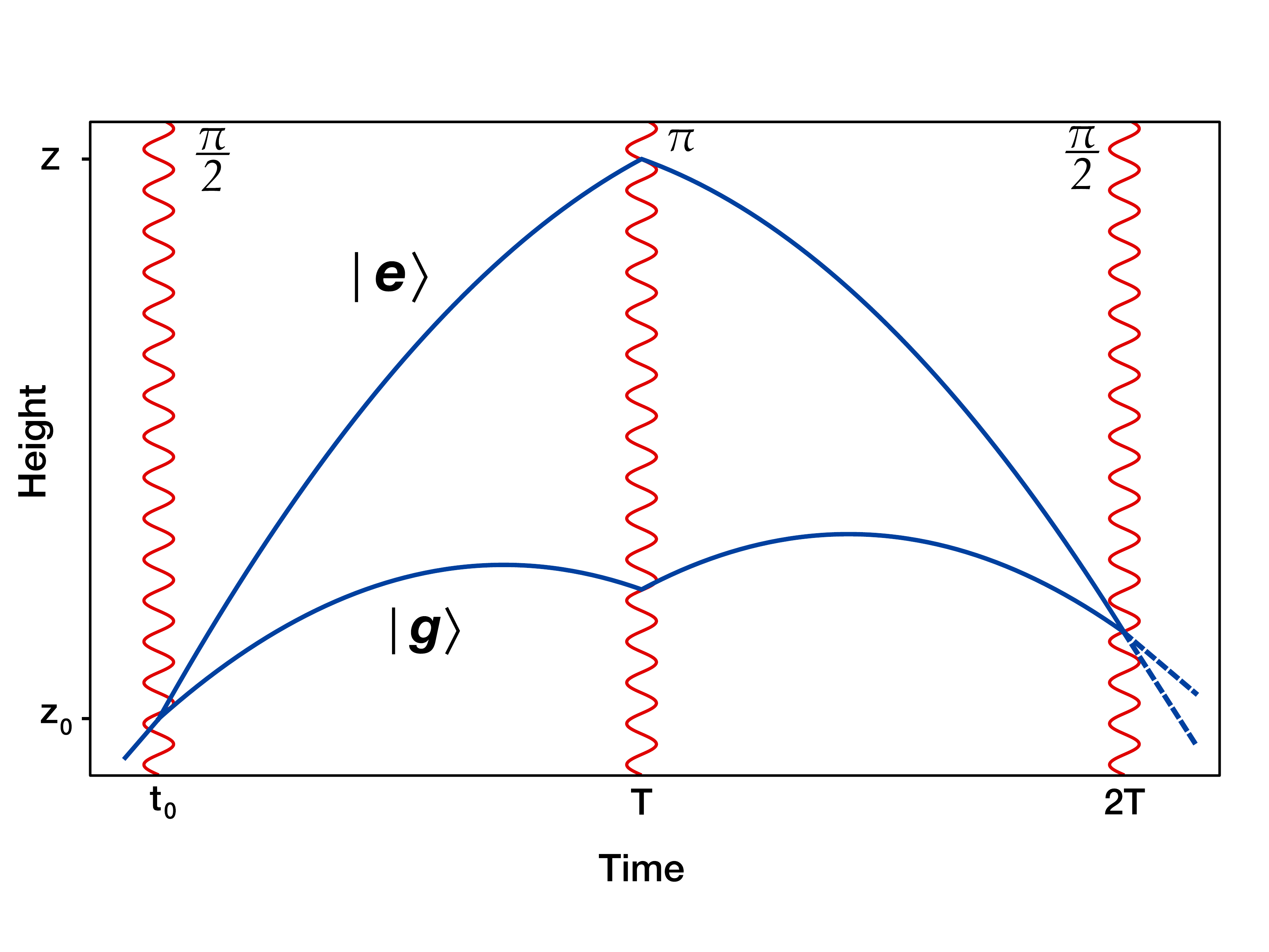}
  \caption{3-pulse Mach-Zehnder light-pulse atom interferometer. Beamsplitter laser pulses are represented as $\pi/2$ and mirror laser pulses as $\pi$. Shown are the respective paths of the ground state and excited state of the initial atom cloud center of mass. The final dashed lines represent the output ports after interference.}
  \label{fig:ai-diamond}
\end{figure}

Over the past several decades atom interferometers have evolved from a tightly constrained laboratory based experiment into a deployable field based detector platform. Current experiments include using atom interferometers in space, underground, on moving vehicles, and being dropped from great heights~\cite{Hogan:2011,tino2013atom,schmidt2011mobile,Bongs:2019,Rasel_2020,Hartwig_2015}. In analogy to laser interferometers, atom interferometers use cold atom clouds in place of laser beams and different spatial trajectories of atom wavepackets as the differing optical paths. The Mach-Zehnder configuration is one of the most common, implementing a three laser pulse sequence while the atom cloud is in free-fall. The first pulse acts as a beam splitter to separate the ground and excited states of the atom cloud; followed by a mirror pulse that redirects the paths of the ground and excited states back towards each other; and finally another beam splitter pulse ensures the trajectories of the two states overlap and the atoms interfere. \Cref{fig:ai-diamond} shows an illustrative example of the center of mass trajectories of both paths of the atom clouds. This all happens in vacuum and the final state of the atoms is interrogated by imaging the atoms using resonant scattering~\cite{tino2013atom}. 

The measurable science signal is the phase difference between the two arms of the atom interferometer. Current atom interferometry experiments have shown the exceptional sensitivity of making fundamental physics measurements ~\cite{asenbaum:2020,Morel:2020,Rosi_2014,Williams_2016,Parker_2018,Kovachy:2015,Hamilton_2015}. Many atom interferometry experiments, such as the MIGA experiment~\cite{Canuel_2018}, the Bremen drop tower~\cite{Muntinga:2013}, ZAIGA~\cite{Zhan_2019}, and the AION experiment~\cite{Badurina_2020} have begun pushing the limits on the terrestrial length and time scales that these detectors can be operated under. In addition, long term vision experiments such as ELGAR~\cite{canuel2020,canuel2020elgar} and AEDGE~\cite{AEDGE:2020} are being developed with international networking in mind.

The Matter-wave Atomic Gradiometer Interferometric Sensor (MAGIS-100)~\cite{abe2021matter} is a large baseline ($\sim$ 100~m) atomic interferometer being built at Fermilab with three strontium based atom sources capable of generating three concurrent atom interferometers along the baseline. The installation environment can contribute to noise backgrounds in the phase difference measurement. For this reason it is crucial to have a preliminary characterization of the site at Fermilab to better understand the noise that will arise from temperature and pressure fluctuations, and seismic activity.

MAGIS-100 aims to be a next generation probe for dark matter (DM), quantum effects, and gravitational waves (GWs) in the mid-band frequency range of 0.1--10~Hz~\cite{abe2021matter,Coleman:2018ozp,Hogan_2016}. The detector uses advances in atomic interferometry while simultaneously pushing the limits of baseline length, time duration, and large momentum transfer (LMT)~\cite{rudolph2020,abend2017,mazzoni2015large,Close:2013,chiow2011,clade2009,muller2009,muller2008,mcguirk2000}. We will also be using the MAGIS apparatus as a stepping stone and pathfinder for examining the effects of the environment at much larger scales and in less controlled sites such as access and mining shafts. All of this effort will be put towards development of future terrestrial extreme baseline atom interferometers at the 1~km scale. Future detectors will most likely be run in even less controlled areas (i.e. mining shafts) where the ability to \textit{in situ} measure the environment and mitigate the noise will be especially important.

\section{Motivation}
\label{sec:motivation}

In order to precisely manipulate the atoms, by laser cooling and atom optics, the environmental effects on the atom interferometer phase needs to be well-measured and analyzed. This includes understanding the active dynamics of temperature, and local seismic activity in and around the shaft. The temperature fluctuations play a role in both the effects on the materials of the vacuum tube and atom source chambers, such as thermal expansion of the metals leading to misalignments, as well as creating the need for a temperature controlled room for the laser systems, which will aim to keep temperatures stable to $\pm\SI{0.3}{\celsius}$, and enclosures for the atom sources and electronics for creating the atom clouds with a temperature stability control of $\pm\SI{1}{\celsius}$. Seismic activity plays two roles in affecting the atom phase. There are two effects caused by direct vibrations and an indirect effect of the ground motion. The first direct effect is that of vibrations of the lasers used to interact with the atoms causing laser phase noise to be imprinted on the atoms. The second direct effect is vibrations of the atom sources leading to fluctuations in the initial atom cloud kinematics that couple to gravity gradients and laser wavefront perturbations~\cite{abe2021matter}. The indirect effect is a fluctuation of the local gravitational potential sourced by mass density fluctuations of the ground known as gravity gradient noise (GGN).

For the experimental sensitivities that we hope to achieve with MAGIS-100, understanding and being able to map out GGN is important because it sets a noise floor for terrestrial measurements~\cite{Saulson:1984}. GGN arises as a secondary effect caused by ground motion and atmospheric pressure fluctuations, and alters the trajectories of the atoms while in free-fall. First, it is an important background that will limit our sensitivity in the mid-band especially when we move to frequencies lower than \SI{1}{\hertz}. Second, it is a theorized effect that has not yet been measured in this frequency range.

From seismology we know that mass density perturbations -- these include earthquakes, slip faults, or other irregular density  fluctuations -- cause waves to propagate through the surface of the Earth. We classify these waves as primary and secondary waves. The former is made up of longitudinal waves while the latter is a transverse wave. The third relevant type of wave is the Rayleigh wave, which contains components of both primary and secondary waves. A common place where these appear is as waves along the surfaces of lakes and rivers. These waves are also important because they are surface waves that can exist along interfaces between different media. Included in these interfaces is the surface of the Earth and the atmosphere~\cite{Viktorov:2013}. The direct effect of these seismic waves is mechanical vibration of the apparatus and lasers. We also see a perturbation in the gravity field's potential. Since our masses are measured in a differential configuration~\cite{Graham:2013}, the vibration associated with the lasers used to split the atom clouds at the top and bottom of the baseline is subtracted out to a high degree.

\section{Site overview}
\label{sec:site-overview}

\begin{figure}
  \centering
  {\includegraphics[width=.7\textwidth]{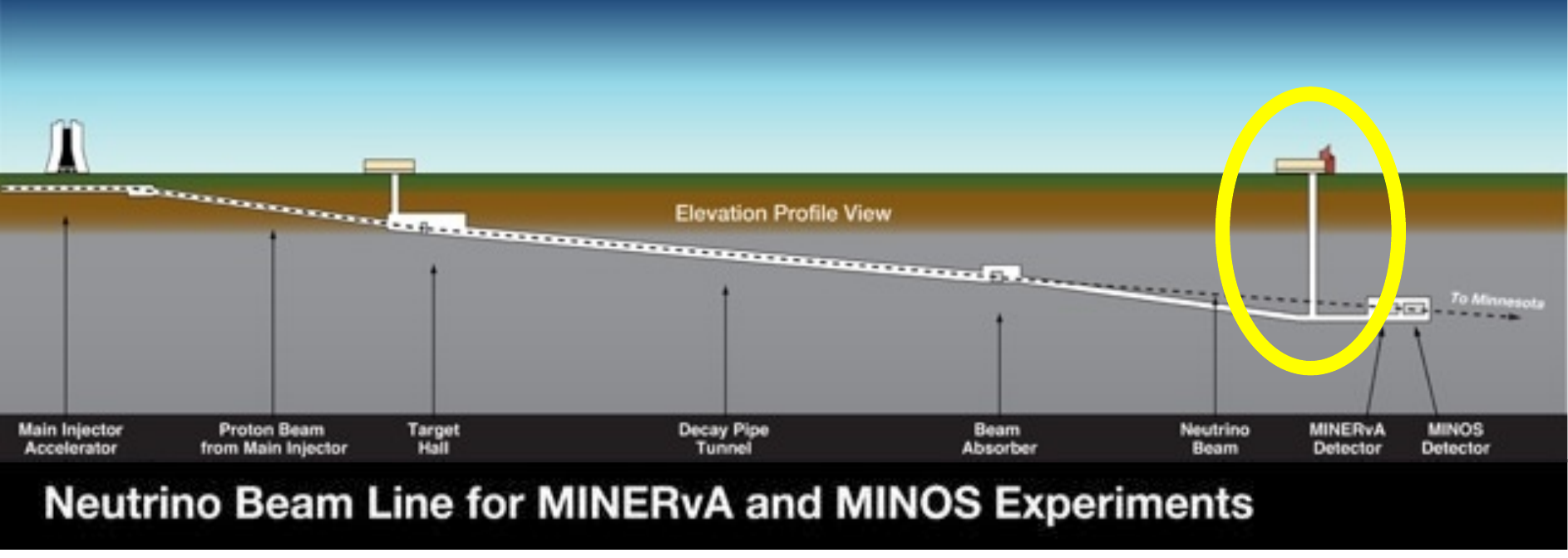}}
  {\includegraphics[height=5cm]{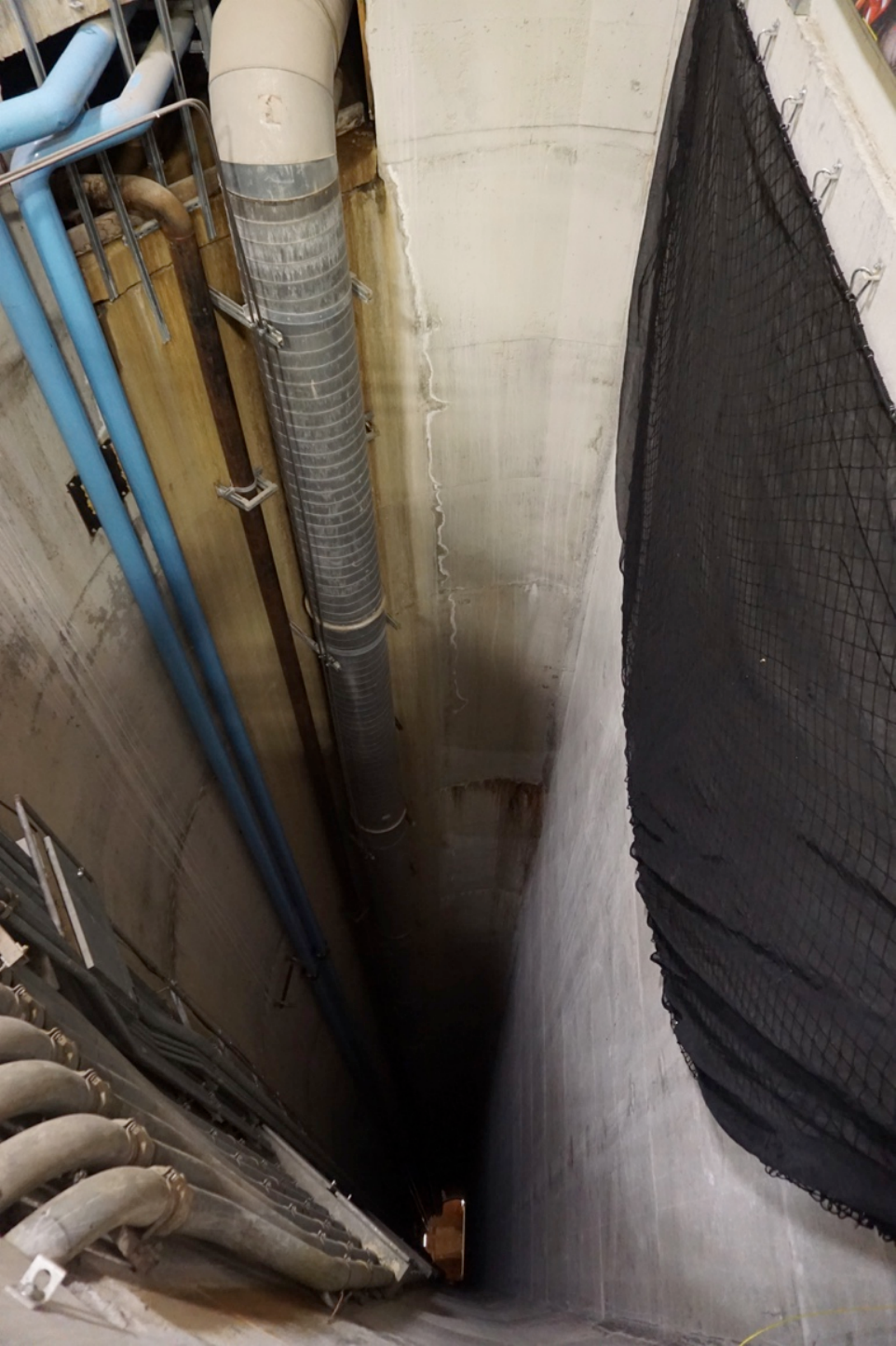}}
  \caption{Location of the MAGIS-100 detector on Fermilab campus. Image of the shaft environment.}
  \label{fig:site-schematic}
\end{figure}

The installation location for MAGIS-100 is a 100~m deep access shaft used primarily for moving underground-experiment equipment at Fermilab, shown in figure \ref{fig:site-schematic}. The shaft is housed in a building with an access elevator adjacent and a rolling door for entry and exit of equipment and materials. MAGIS-100 has two main locations within the MINOS service building. These are the laser room required to house all of the lasers for generating and manipulating the atoms and a vacuum tube with the atom sources attached installed along the vertical wall of the shaft. Many forms of noise mitigation will be directly implemented by proper construction of the laser room and civil engineering for the installation of the vacuum tube. Measurements of the other noise sources and sensitivity limiting effects will be discussed and analyzed in sections~\ref{sec:methods} and \ref{sec:results}.

\section{Measurement methods and modeling}
\label{sec:methods}

\subsection{Temperature and humidity measurement methods}

For temperature, humidity, and dew point measurements we use an Omega OM-DVTH Temperature/Humidity data logger. They are compact and battery run which allows for placing them in multiple locations around the MAGIS-100 site. We placed one logger at the top of the shaft and another underground near the bottom of the shaft. Data was then sampled every minute for a period from January 2019 to June 2019. The temperature accuracy of the logger is $\pm\SI{0.5}{\degree C}$ with a resolution of \SI{0.005}{\degree C}. Relative humidity (RH) accuracy and resolution are $\pm 2\%$ and $0.01\%$, respectively.

\subsection{Seismic measurement methods}

A Trimble RefTek 151B-120 Observer seismometer was installed in two locations to measure ground motion. Our primary focus was to measure the vertical displacement and acceleration of the ground, which is driven by Rayleigh waves along the surface. In order to measure the ambient seismic noise we needed a low self-noise seismometer. As a preliminary step we collected data from the surface of the shaft and underground near the lower exit of the shaft. The environment underground is much quieter so the two locations allow us to look for various vibrational noise sources that could be contaminating the signal at the surface. The seismometer has a self-noise lower than the New Low Noise Model (NLNM) from 0.0167~Hz--10Hz. The NLNM as well as the New High Noise Model (NHNM) were established using a global network of seismograph stations by the United States Geological Survey (USGS) to set a general model for quiet and noisy environment backgrounds. This makes it ideal for the modeling we performed. Data was then collected for 2 months above ground and 9 months underground. Data collection was taken at a sampling rate of 50 samples per second above ground and 40 samples per second underground. The time traces were sampled into hour long segments and power spectra of the data were then constructed. We followed the same methods used to construct the NLNM and NHNM power spectral densities (PSD) from the recorded seismic data~\cite{Peterson:1993,mcnamara2006seismic}. These hourly power spectra could then be averaged and binned in a 2D histogram for the period of observation and the mean values analyzed.

Vertical displacement power spectra were then used to model an inferred GGN strain spectrum. Since seismic isolation and Newtonian noise are also systematics in laser interferometers, we followed previous methods employed at LIGO to model GGN~\cite{Harms:2015}. We approximate the ambient GGN strain spectrum by measuring the vertical seismic motion, constructing the associated displacement power spectral densities, and evaluating the model for the resulting spectrum. We only consider vertical motion because the Rayleigh surface waves, which compose most of the seismic noise we are interested in, have coupled horizontal and vertical motion with respect to the interface surface. Thus, only one direction of motion is needed to fully describe the Rayleigh wave~\cite{Viktorov:2013}.

The perturbation on the acceleration of a test mass can be written as
\begin{equation}
  \label{eq:1}
  \var{a_{z}}(\vec{r}_{0},t) = -2\pi G \rho_{0} \mathrm{e}^{-zk_{\ell}}\mathrm{e}^{i(\vec{k}_{\ell}\vdot\vec{\ell}_{0}-\omega t)}\gamma(\nu)\xi_{z},
\end{equation}
where $G$ is Newton's constant, $\rho_{0}$ is the average ground density, $\gamma (\nu) \approx 0.27$ is a ground material term dependent on the Poisson ratio with a typical range of 0.2--0.35~\cite{Zandt:1995}, $\nu$ is the Poisson ratio of the ground, $z$ is the depth from the surface, and $\vec{\ell}_{0}$ is the projection of the position vector on the surface. We assume a seismic Rayleigh wave with wave vector $\vec{k}_{\ell}$ projected onto the surface, frequency $\omega$, and displacement field $\xi_{z}$~\cite{Harms:2015}. For simplicity this form does not include extra terms from underground bulk compression waves. This scenario is realized for an atomic cloud as our test mass falling vertically down a shaft. The acceleration difference of the atom cloud measured at two different vertical locations will influence the relative phase during an atom interferometric sequence, and will be measured as a relative phase error between the two positions in a gradiometer configuration. For two atom interferometers separated by a baseline distance $L=h_{2}-h_{1}$ this leads to a difference in the perturbed acceleration given by
\begin{align}
  \label{eq:2}
  \var{a_{z}(h_{2})}-\var{a_{z}(h_{1})} &= 2\pi G\rho_{0}\gamma(\nu)\xi_{z}\qty(\mathrm{e}^{-k_{\ell}h_{1}}-\mathrm{e}^{-k_{\ell}h_{2}}).
\end{align}
There is a cross-over depending on if the baseline length is longer or shorter than the wavelength of the Rayleigh wave. There are two regimes for low and high frequencies, $f < \frac{c_{\mathrm{R}}}{2\pi L}$ and $f > \frac{c_{\mathrm{R}}}{2\pi L}$ where $c_{\mathrm{R}}$ is the ambient Rayleigh wave speed, that need to be considered. The values of $c_{\mathrm{R}}$ range between 3 -- 5~km/s in solid metal and rock and between 200 -- 400~m/s in shallow Earth. For long wavelength (low frequency) Rayleigh waves, the GGN strain $h_{\mathrm{GGN}}=\expval{\delta x_{z}} /L$ where $\expval{\delta x_{z}}$ is the displacement perturbation spectrum, can be expanded in terms of $k_{\ell}L \ll 1$ as
\begin{equation}
  \begin{split}
    \label{eq:3}
    h_{GGN} &= \frac{G\rho_{0}\gamma (\nu)k_{\ell}}{2\pi f^{2}}\expval{\xi_{z}}\\
    &= \frac{G\rho_{0}\gamma (\nu)}{f c_{R}}\expval{\xi_{z}}.
  \end{split}
\end{equation}

In the short wavelength (high frequency) regime the strain noise is
\begin{equation}
  \label{eq:4}
  h_{GGN} = \frac{G\rho_{0}\gamma (\nu)e^{-k_{\ell}L}}{2\pi f^{2}L}\expval{\xi_{z}}.
\end{equation}

Using recorded seismic vertical displacement data $\xi_{z}$ from the seismometer we use the above model to estimate an inferred GGN strain spectrum for MAGIS-100. 

\section{Results}
\label{sec:results}

\subsection{Temperature and humidity}

\begin{figure}
  \centering
  \begin{minipage}[htb!]{0.5\textwidth}
    \includegraphics[width=0.95\textwidth]{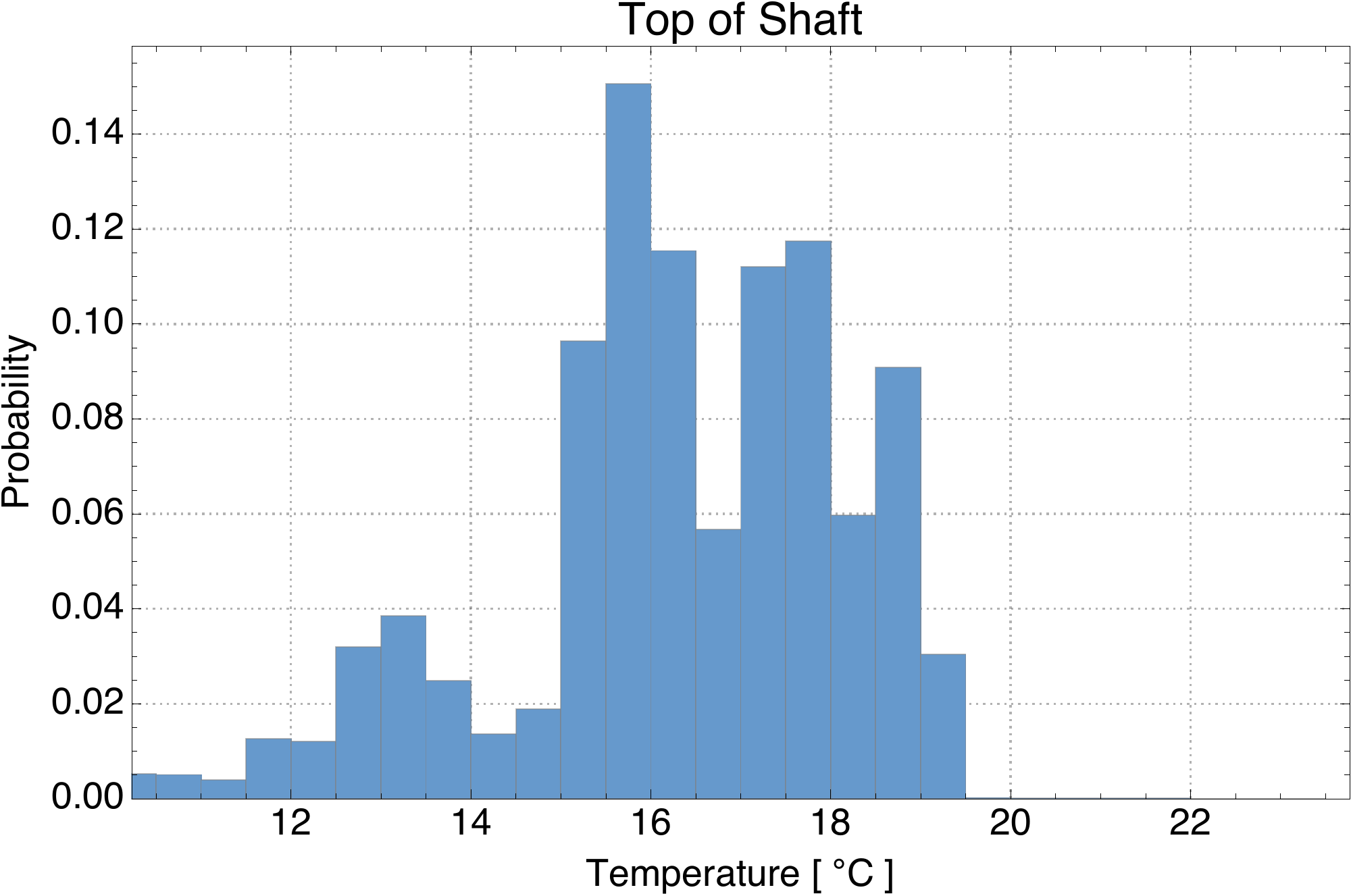}
  \end{minipage}%
  \begin{minipage}[htb!]{0.5\textwidth}
    \includegraphics[width=0.95\textwidth]{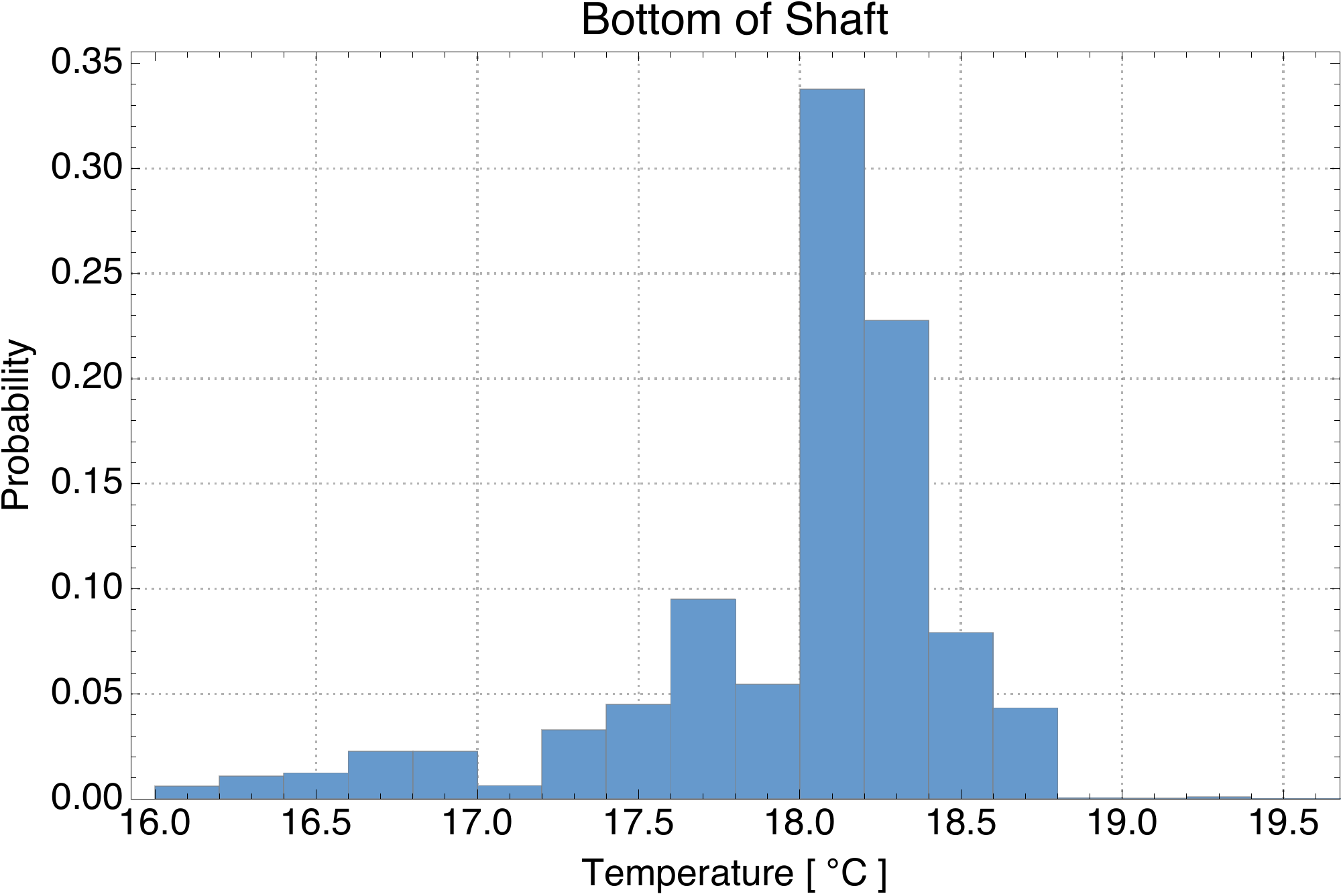}
  \end{minipage}%
  \caption{Temperature recorded at the top of the shaft in the MINOS service building and at the bottom of the shaft.}
  \label{fig:temp-hist}
\end{figure}

For the period between late January and early June analysis of the temperature readings show a tightly constrained distribution underground at the bottom of the shaft and a slightly larger spread of temperatures at the surface. Data from the hottest months of July and August were not available as laboratory activities restricted the measurement campaign but will be studied further and continuously monitored during the duration of the experiment. As can be seen in \cref{fig:temp-hist}, above ground there exist two peaks around \SI{15.5}{\celsius} and \SI{17}{\celsius}. This variation implies a need to have an extra layer of temperature control at the surface. The MINOS service building that houses the shaft has a large roll-up door that provides large surface area for temperature change when opened. Underground the environment is controlled through air-handling units and the only connections to the surface environment are through the shaft and a smaller pipe near the end of the underground tunnel.

\begin{figure}
  \centering
  \begin{minipage}[htb!]{0.5\textwidth}
    \includegraphics[width=0.95\textwidth]{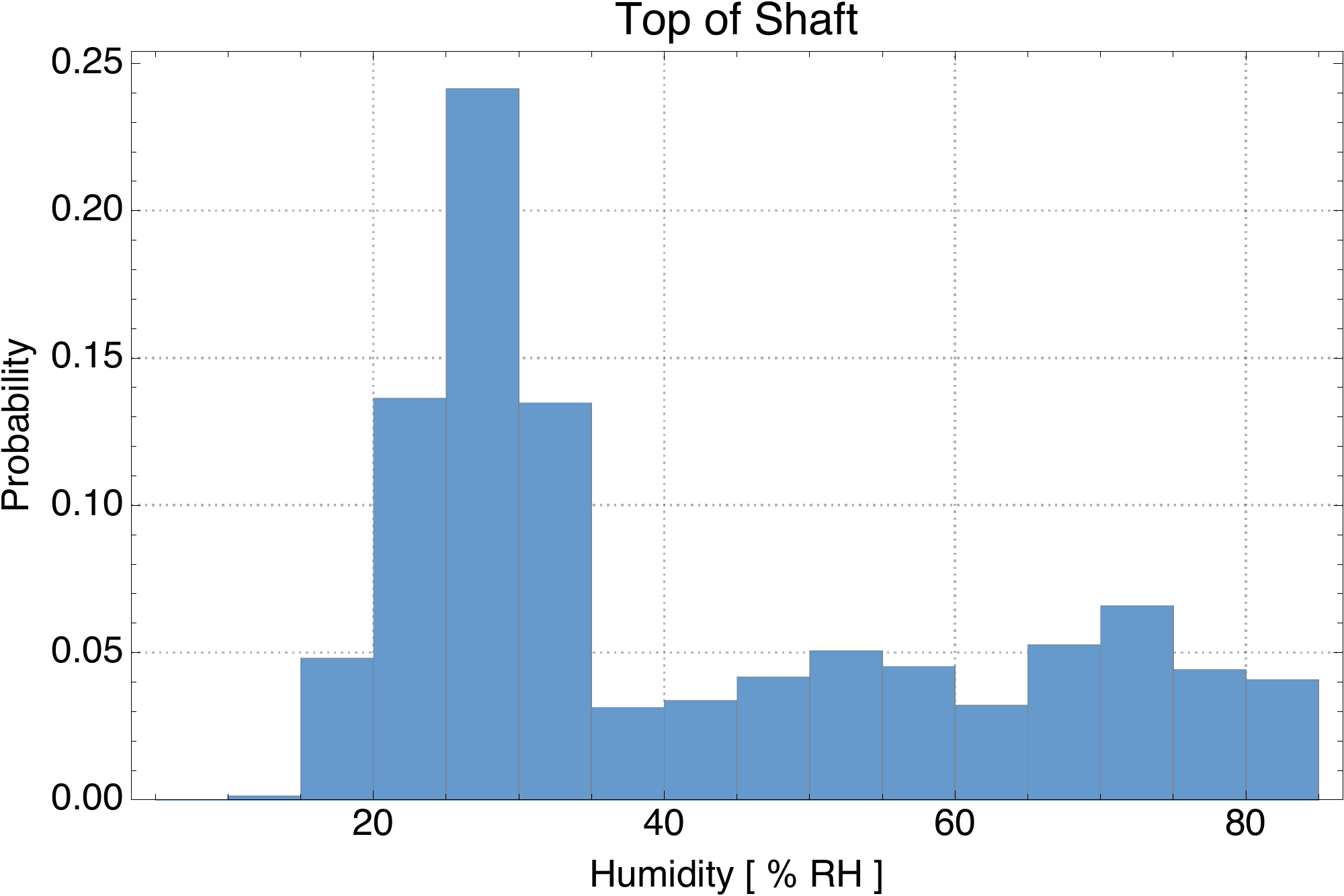}
  \end{minipage}%
  \begin{minipage}[htb!]{0.5\textwidth}
    \includegraphics[width=0.95\textwidth]{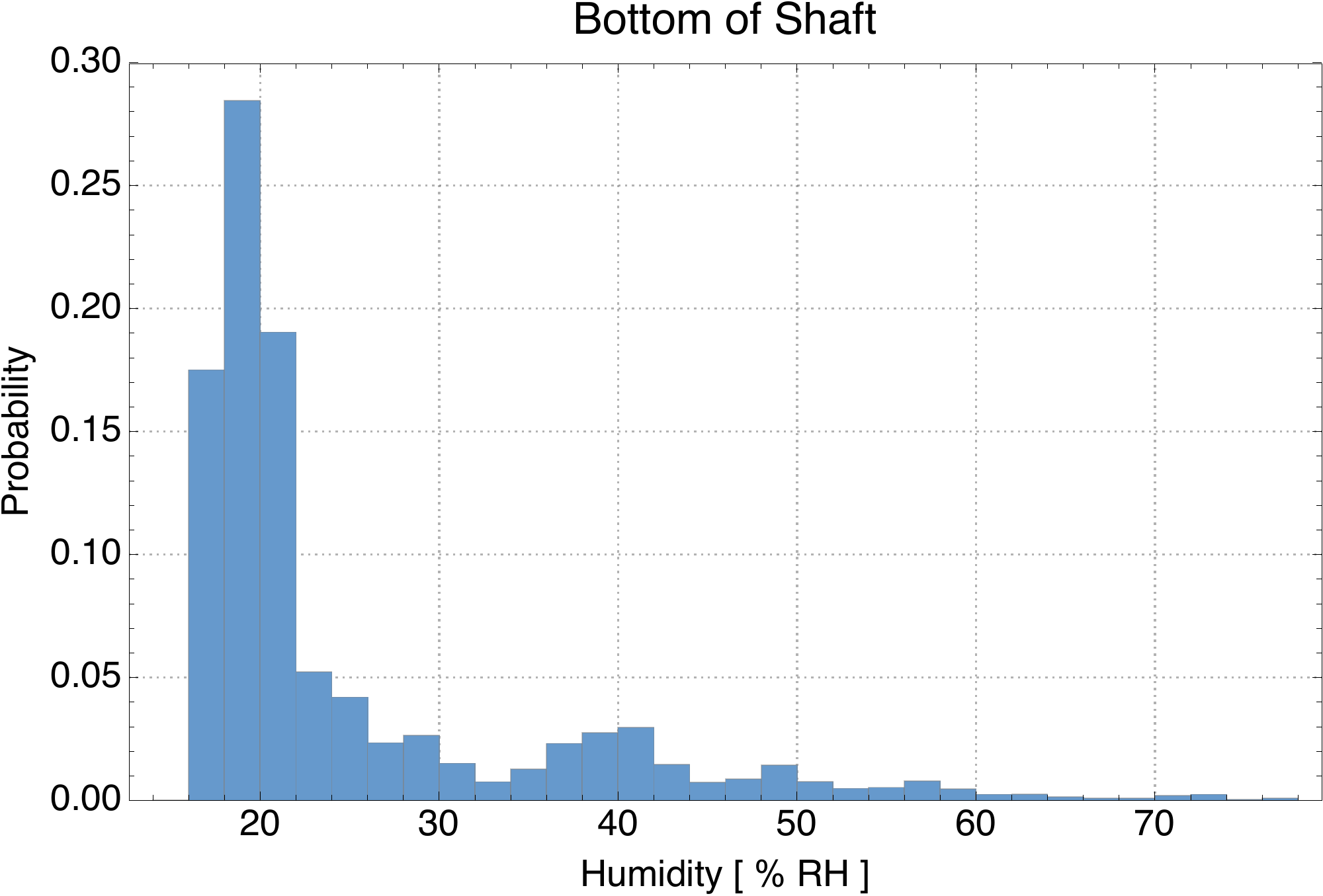}
  \end{minipage}%
  \caption{Humidity recorded at the top of the shaft in the MINOS service building and at the bottom of the shaft.}
  \label{fig:hum-hist}
\end{figure}

\begin{figure}
  \centering
  \begin{minipage}[htb!]{0.5\textwidth}
    \includegraphics[width=0.95\textwidth]{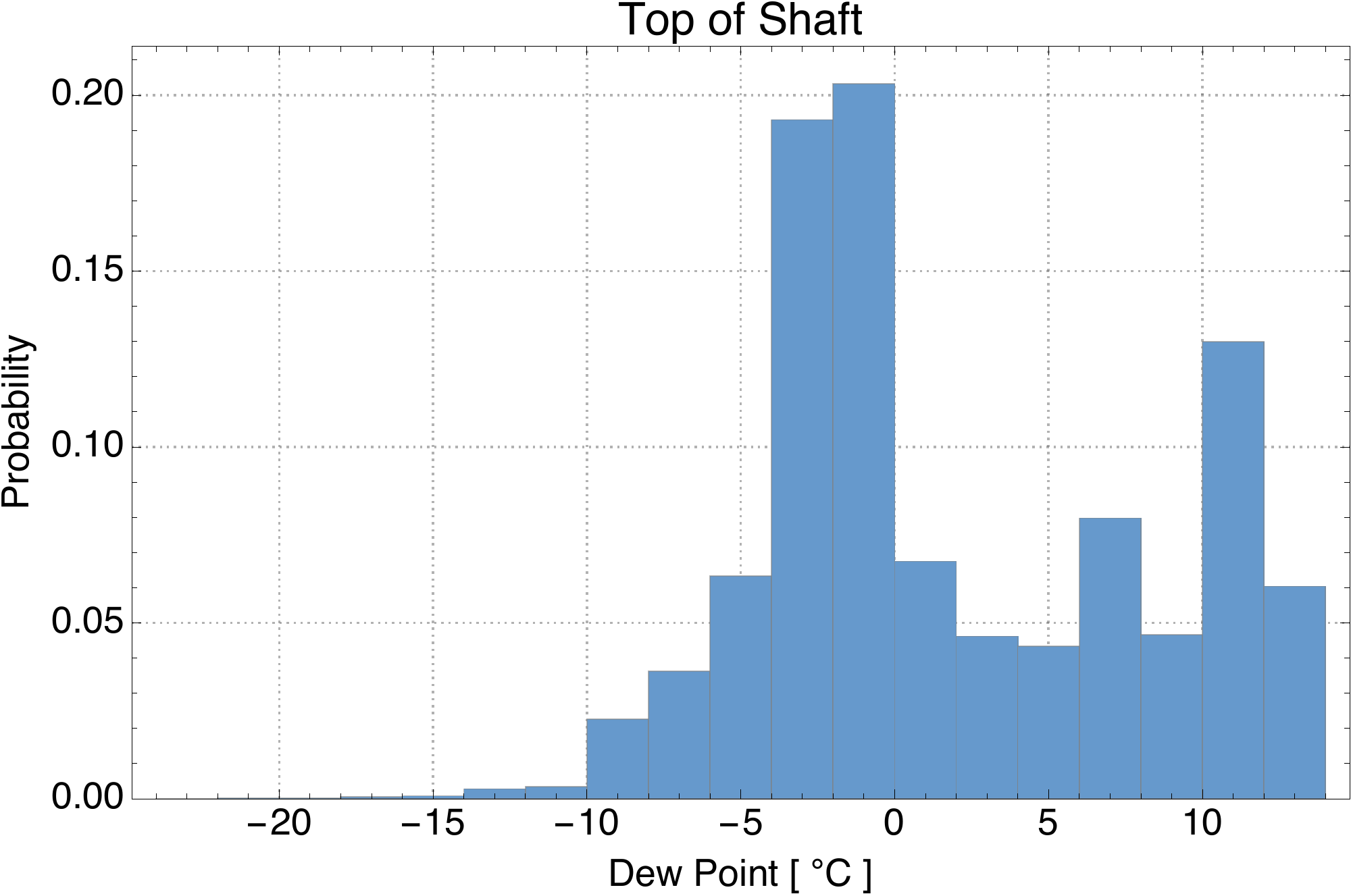}
  \end{minipage}%
  \begin{minipage}[htb!]{0.5\textwidth}
    \includegraphics[width=0.95\textwidth]{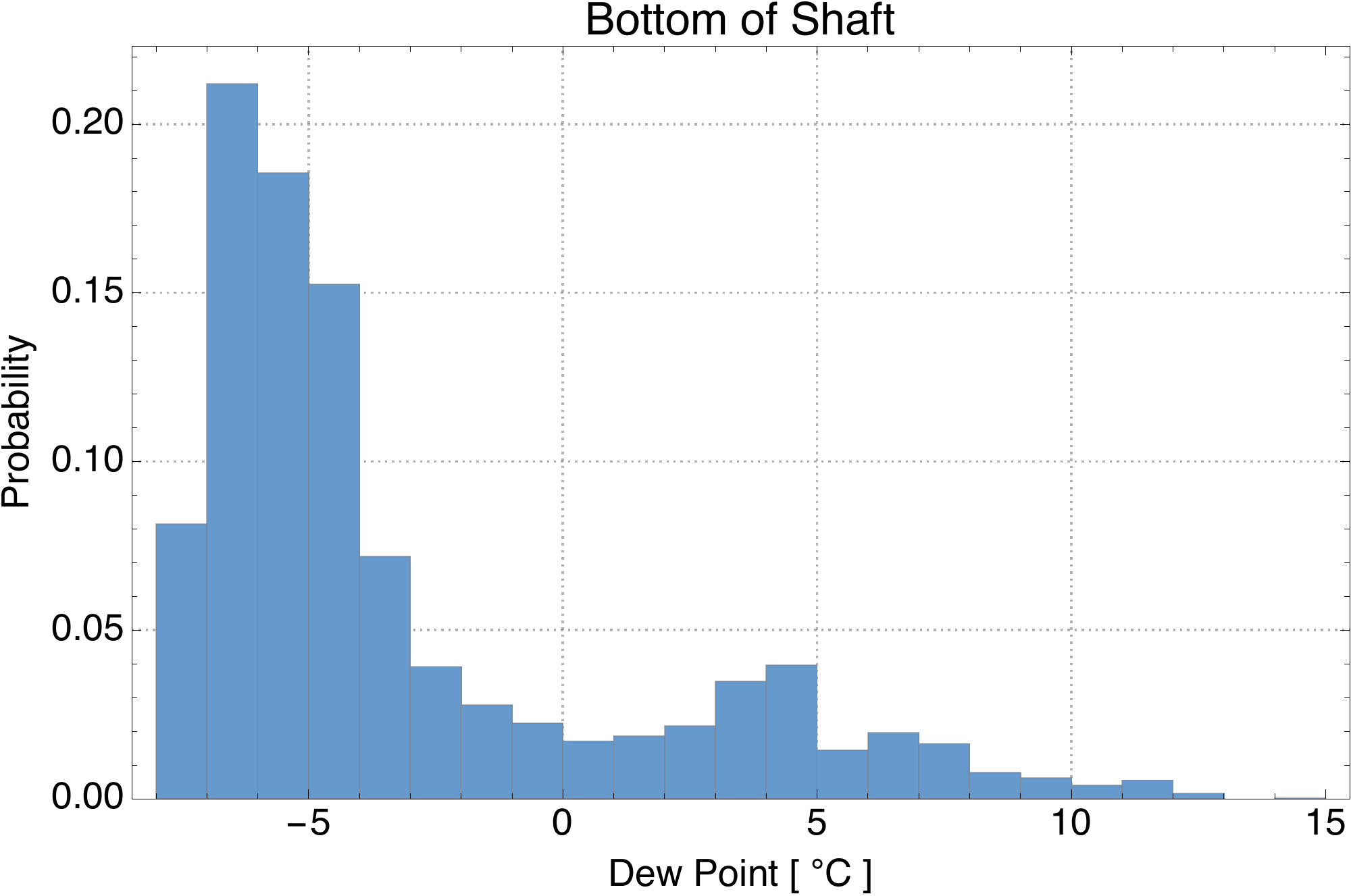}
  \end{minipage}%
  \caption{Dew point recorded at the top of the shaft in the MINOS service building and at the bottom of the shaft.}
  \label{fig:dew-hist}
\end{figure}

Humidity and Dew point follow the same pattern having a narrow distribution underground and a broader distribution above ground shown in \cref{fig:hum-hist,fig:dew-hist}. This is again caused by the control of the underground air and the variability of the temperature caused by the opening and closing of the entry way to the MINOS building. The humidity at the surface level has a peak between 25--35\%RH. Below ground the peak humidity remains between 18--20\%RH. Above ground the humidity fluctuates with the seasons between a large range of roughly 20--80\%RH. The dew point underground has a peak between \num{-8}--\SI{-3}{\celsius}, while at the surface there are two main peaks in the ranges \num{-3}--\SI{0}{\celsius} and \num{10}--\SI{11}{\celsius}, respectively.

\begin{figure}
  \centering
  \begin{minipage}[htbp!]{0.45\textwidth}
    \includegraphics[width=\textwidth]{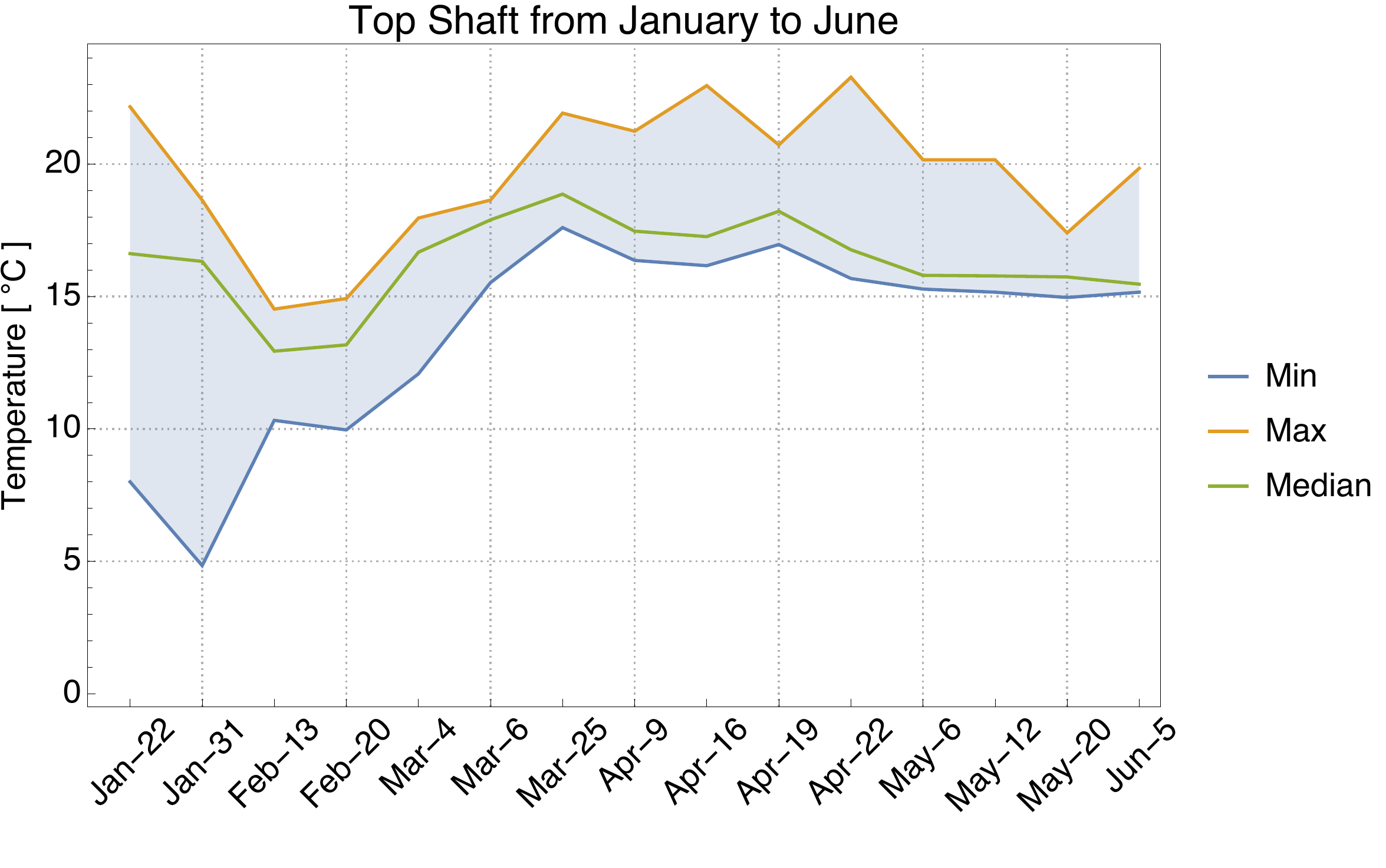}
  \end{minipage}%
  \hspace{1em}
  \begin{minipage}[htbp!]{0.45\textwidth}
    \includegraphics[width=\textwidth]{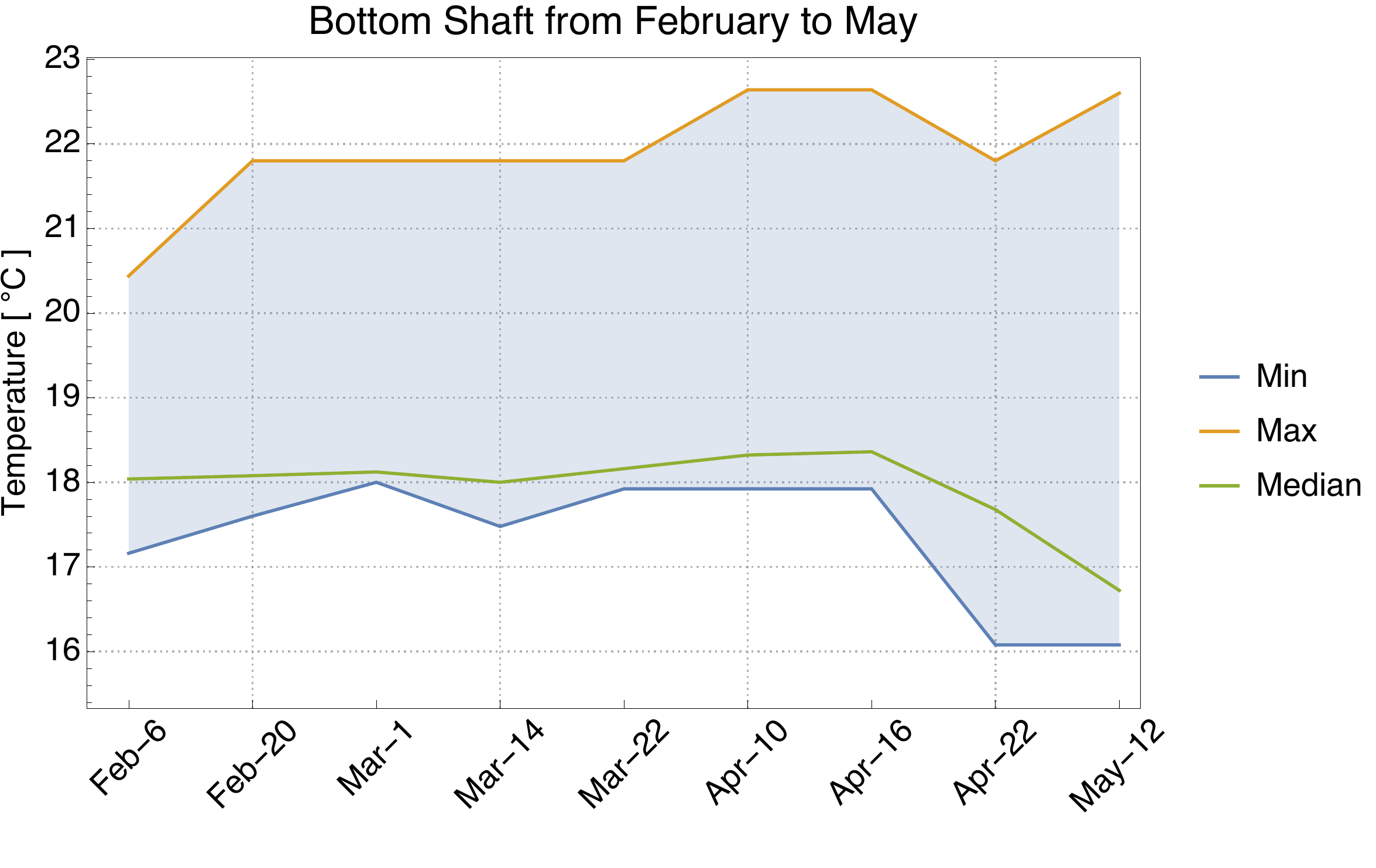}
  \end{minipage}%
  \caption{Minimum, Maximum, and Median values of weekly temperature at the top of the MAGIS shaft and the bottom of the shaft over a 5 month period and a 3 month period, respectively.}
  \label{fig:minmaxplots}
\end{figure}

\begin{figure}
  \centering
  \begin{minipage}[htb!]{0.4\textwidth}
    \includegraphics[width=\textwidth]{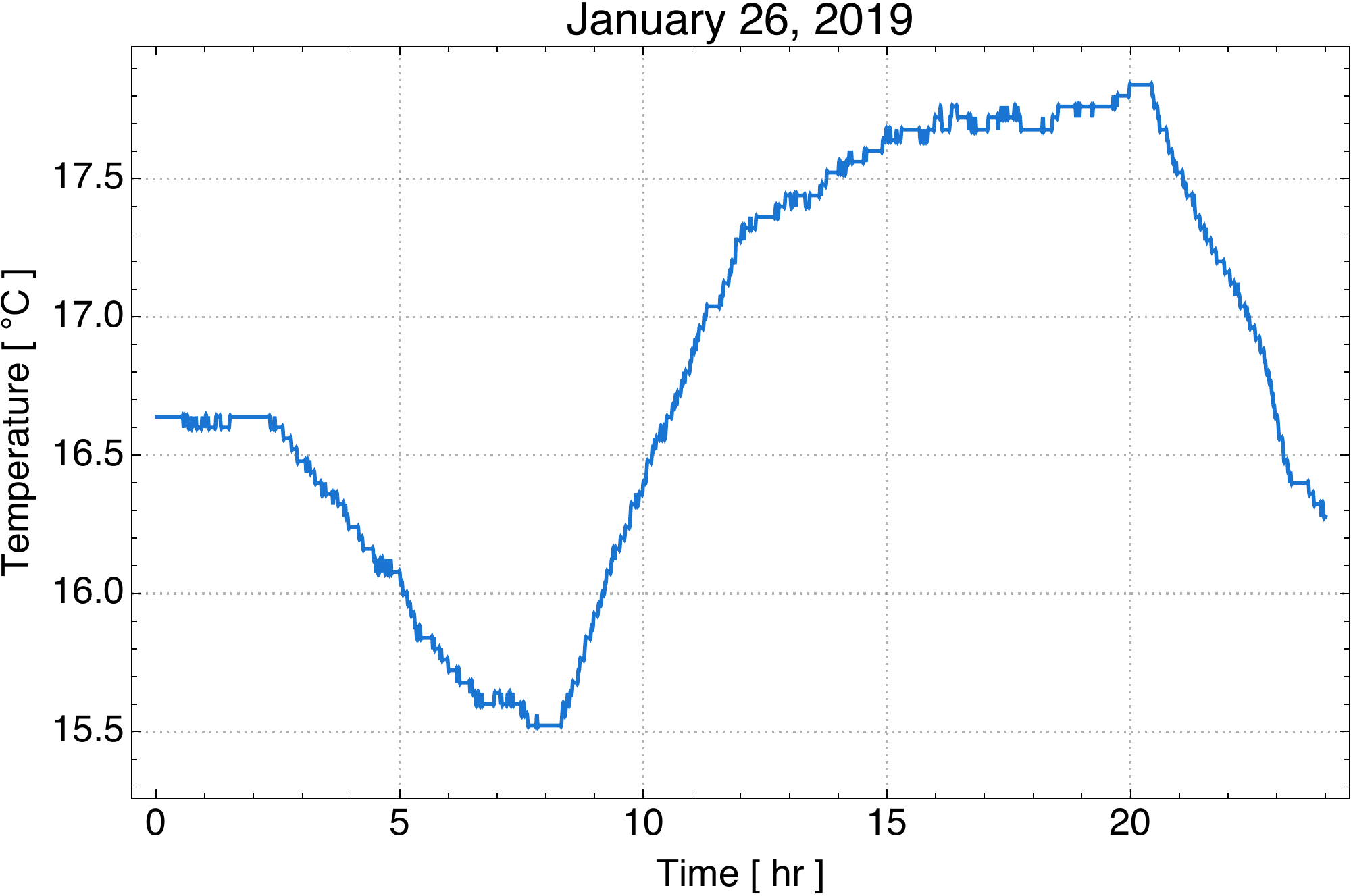}
  \end{minipage}%
  \hspace{2em}
  \begin{minipage}[htb!]{0.4\textwidth}
    \includegraphics[width=\textwidth]{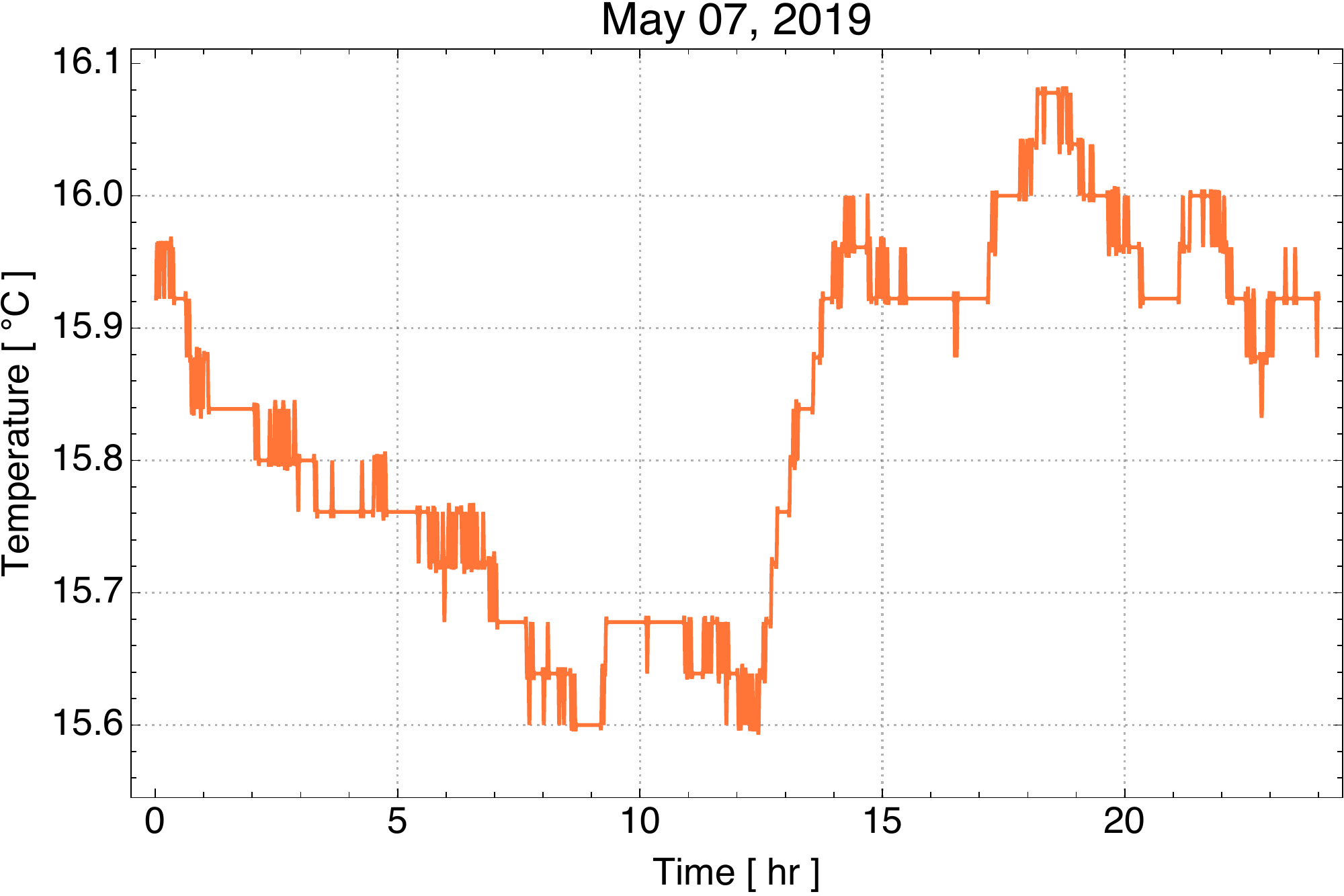}
  \end{minipage}%
  \caption{Typical daily temperature fluctuation at the surface for January 26 and May 07.}
  \label{fig:daytemp}
\end{figure}

Figure \ref{fig:minmaxplots} shows median temperature readings at the surface remain bound between \num{13}--\SI{19}{\celsius} for the full 5 month period of temperature data, but fluctuate with the seasons. Underground median temperatures are between \num{16.5}--\SI{18.5}{\celsius} and fluctuate slowly with the seasons because of the controlled air-conditioned environment. We also see some time dependent fluctuations of temperature at the surface shown in figure~\ref{fig:daytemp} for average days in January and May.

\subsection{Seismic and gravity gradient noise}

\begin{figure}
  \centering
  \includegraphics[width=\textwidth]{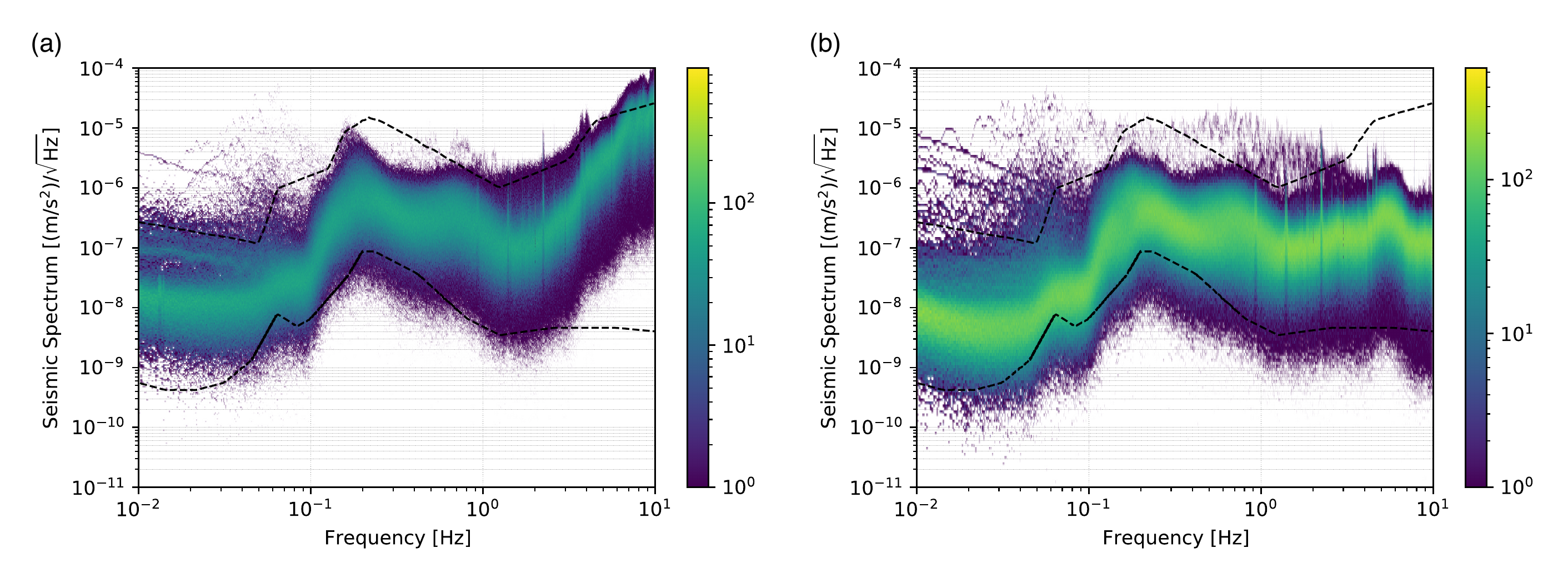}
  \caption{Seismic amplitude spectra for the surface of the shaft (a) and underground near the exit of the shaft (b). The color bar represents the histogram counts for the amplitude for each frequency bin. Also plotted are the NLNM (lower black line) and the NHNM (upper black line).} 
  \label{fig:seismic-spectra}
\end{figure}

Seismic data binned by recorded frequency is shown in~\cref{fig:seismic-spectra}. The vertical acceleration seismic amplitude spectra above and below ground are between the NLNM/NHNM (depicted on the histograms as the lower and upper dashed black lines, respectively). Data was taken from December 18, 2018 to February 25, 2019 for surface measurements and from February 25, 2019 to May 30, 2019 for the underground measurements. Below \SI{1}{Hz} the Earth's primary and secondary microseism peaks can be seen. They are a global background phenomena sourced by ocean waves interacting with the ocean floor and excite the Earth's natural modes which can be seen at any point on the surface. They are also a good indicator of the seismometer's sensitivity~\cite{Nishida:2013}. Higher frequency content in the amplitude spectrum can be attributed to mechanical vibrations of nearby air-handling units, air compressors, and a ratchet-clank elevator used for accessing the underground facilities.

\begin{figure}
  \centering
  \includegraphics[width=\textwidth]{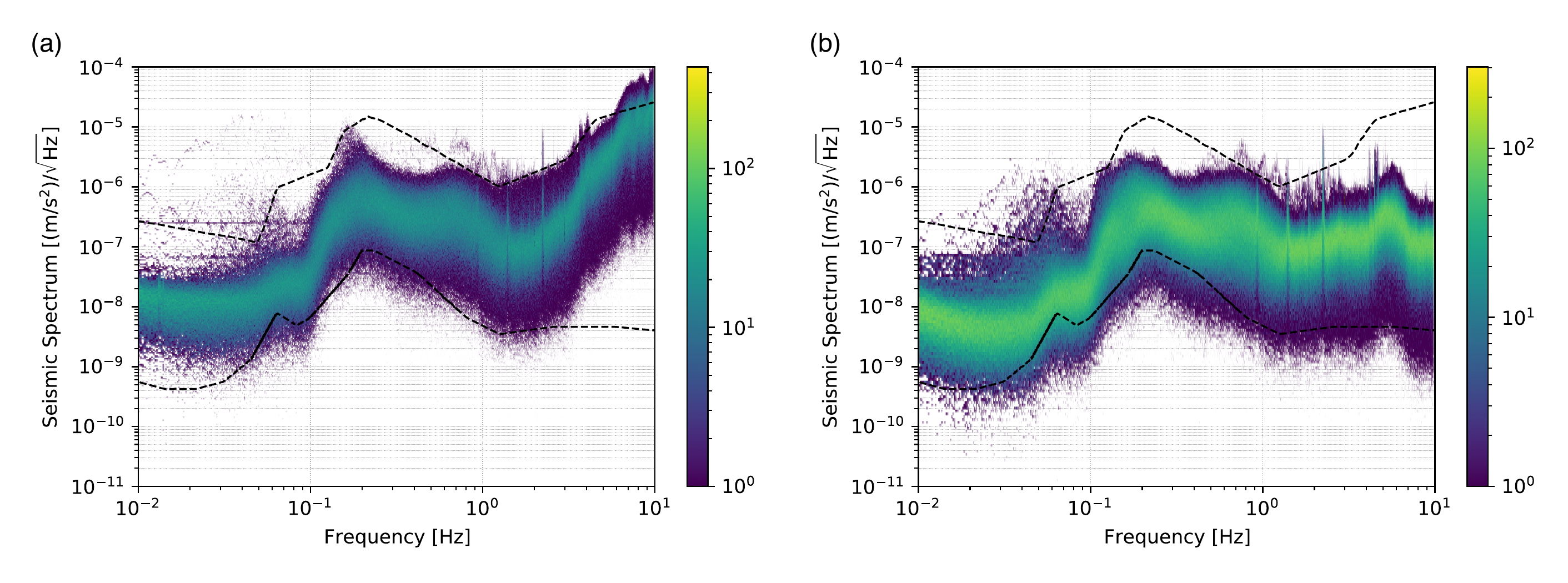}
  \caption{Seismic amplitude spectra for the surface of the shaft (a) and underground near the exit of the shaft (b) between 6~pm and 6~am. The color bar represents the histogram counts for the amplitude for each frequency bin. Also plotted are the NLNM (lower black line) and the NHNM (upper black line).} 
  \label{fig:seismic-spectra-night}
\end{figure}

Comparing surface spectra to underground spectra shows differences in the amplitude of the higher frequency content, which is associated with anthropic generated Newtonian noise. We also observe a decrease in human generated noise at night shown in~\cref{fig:seismic-spectra-night}. Even for the full spectrum the acceleration vibration amplitudes are within acceptable limits for our current design plans including wall mounting optics and support of the vacuum tube.

\begin{figure}
  \centering
  \includegraphics[width=0.6\textwidth]{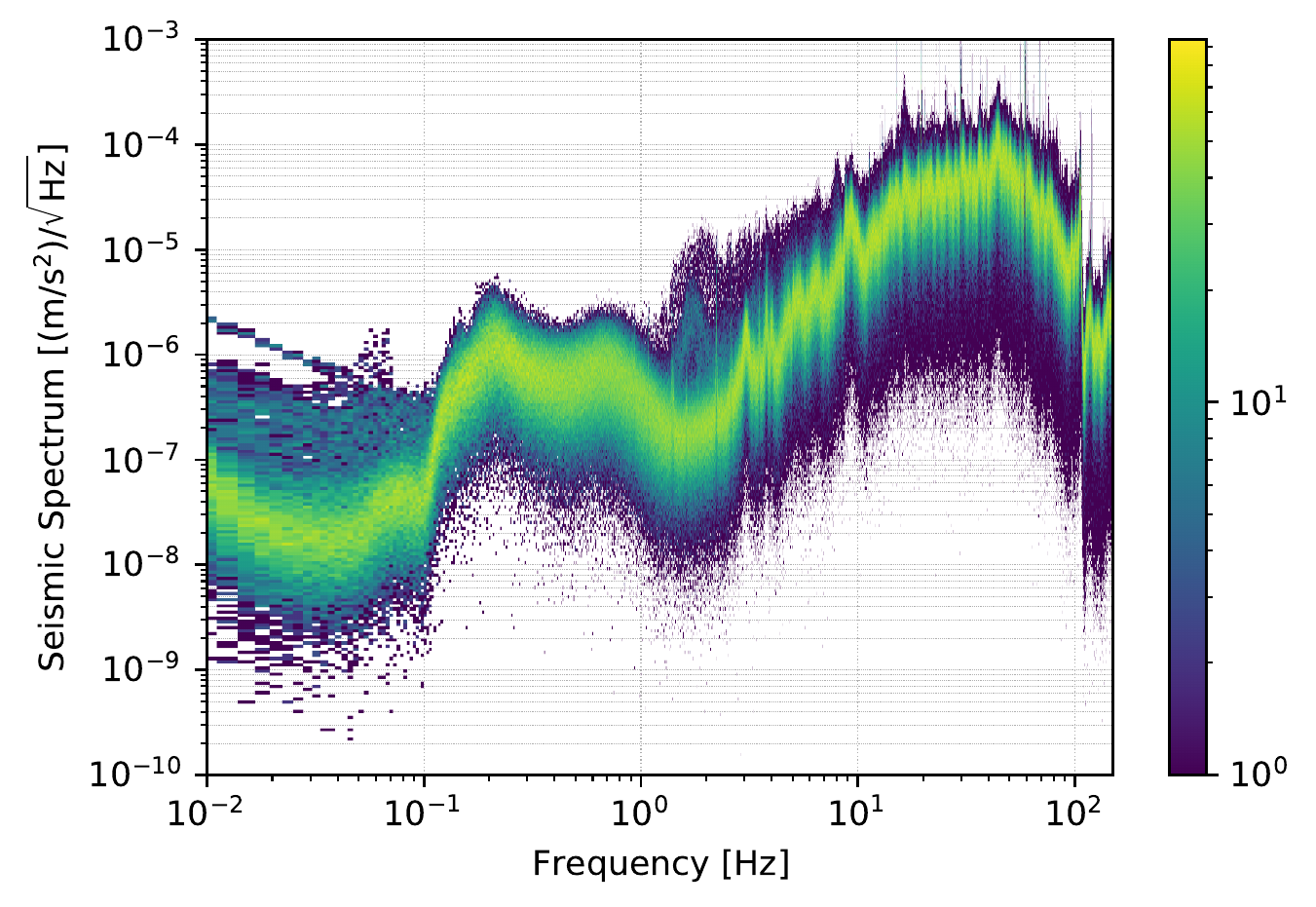}
  \caption{Seismic acceleration spectrum plot at the planned laser table site sampled at \SI{500}{Hz}.}
  \label{fig:seismic-spectra-pedestal}
\end{figure}

In addition to the low frequency seismic acceleration spectrum measurements, we took higher frequency samples in the planned location of the laser system to ensure design specs would meet the required vibration isolation. \Cref{fig:seismic-spectra-pedestal} shows the vibration amplitude power spectrum out to \SI{100}{Hz}.

\begin{figure}
  \centering
  \includegraphics[width=\textwidth]{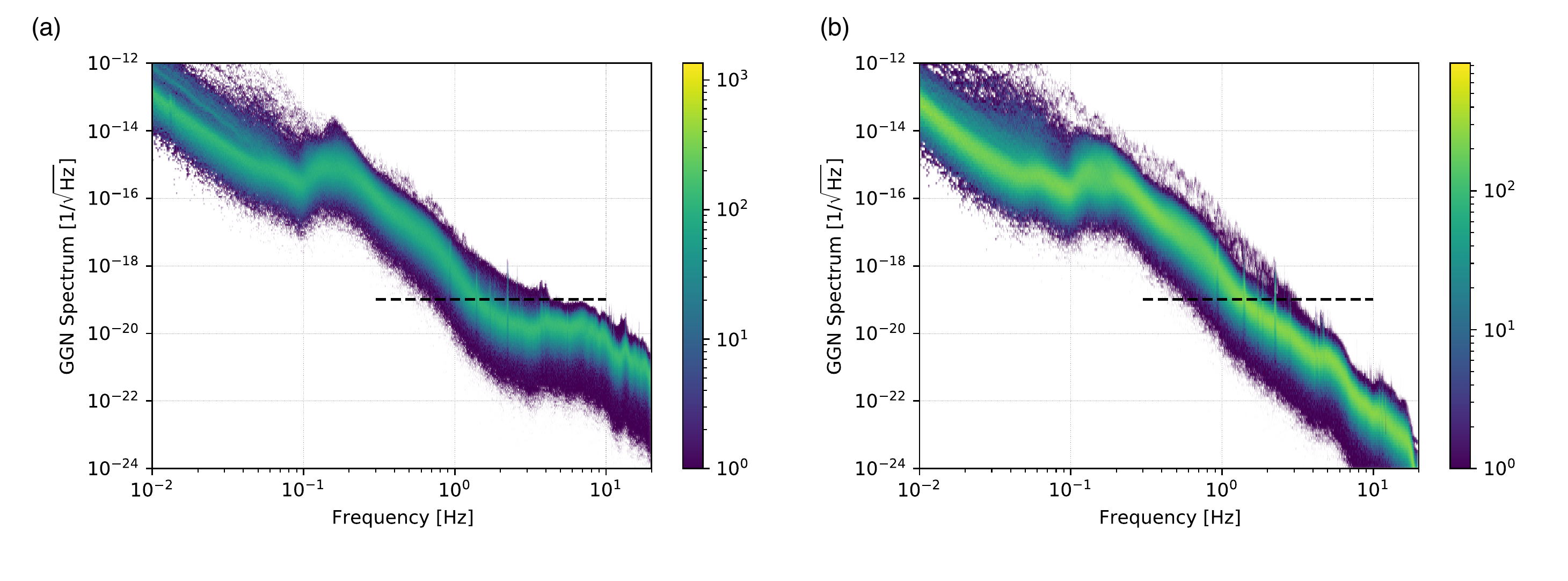}
  \caption{Inferred GGN strain amplitude spectra for the surface of the shaft (a) and underground (b) computed from vertical displacement measurements assuming Rayleigh wave velocity of \SI{300}{m/s}. The dashed black line denotes the expected strain sensitivity after detector advancements~\cite{abe2021matter}. }
  \label{fig:ggn-model}
\end{figure}

Modeled GGN shows levels that will not affect MAGIS-100 until the later stages of the experiment. The modeled GGN, as seen in~\cref{fig:ggn-model}, shows little variation between the surface measurements and the underground measurements as was expected from the analytical model, since the amplitude is exponential with a characteristic length scale of order $k_{\ell} L \sim 1$. The modeled GGN also has a knee frequency of \SI{0.48}{Hz} where the analytical model changes from the high frequency regime to the low frequency regime.

\section{Discussion}
\label{sec:disc}

This early site characterization will guide research and development of the required control and monitoring systems for temperature and seismic vibrations. From the data collected we have been able to set targets for the level of mitigation necessary as well as plans for thermal and vibrational insulation. Preliminary measurements of the site environment show that specifications will be met by incorporating passive and active mitigation, such as temperature controlled enclosures for the atom sources and laser system and vibration isolating tables and mounts for the lasers and optics. Other seismic mitigation will include an active seismometer array. Systematics for the interferometer phase that depend on temperature and seismic field fluctuations include: blackbody radiation shifts, background gas index of refraction, initial cloud kinematics (typical atom numbers range from \num{e6}--\num{e10} atoms per cloud), seismic vibration of laser optics, and GGN.

\subsection{Temperature fluctuation and mitigation}
\label{subsec:temp-disc}

Temperature fluctuations directly impact the atom interferometer phase through blackbody radiation shifts and background gas index of refraction changes. Blackbody radiation can shift the atomic energy levels of strontium resulting in phase noise in the interferometer if the temperature of the vacuum tube fluctuates with a rate inside the target frequency band. With a temperature fluctuation of \SI{1}{\degree C} over a period of 1 hour the noise magnitude is \SI{1e-6}{rad/\sqrt{Hz}}. The effect coming from background index of refraction fluctuations alters the optical path length associated with the baseline of the interferometer. Index of refraction noise leads to false strain signals~\cite{Dimopoulos:2008}. Time dependant fluctuations in the system temperature leads to a strain signal associated with the index of refraction, see~\cite{abe2021matter} for more details.

From the results above it is clear that the temperature fluctuations underground are not of significant concern regarding the previously described systematics. At the surface however, there are daily temperature fluctuations that are aligned with the weather outside of the MINOS service building and the large roll-up door provides a wall of external air to enter all at once. We intend to restrict the opening of the large door during data runs as this will allow control of the largest temperature deviations. The atom sources to be installed down the shaft wall will also be sealed in temperature controlled enclosures. Finally temperature and humidity controls will be incorporated into the laser room design. Long term diagnostics of the temperature, humidity, dew point, and pressure will also be actively measured and analyzed by an environmental DAQ system. These mitigation strategies will reduce the systematic noise in the phase measurements related to blackbody radiation shifts, and background gas index of refraction fluctuations in the vacuum tube~\cite{A.-Sugarbaker:2013,Chris-Overstreet:2018,Susannah-M.-Dickerson:2013}.

\subsection{Seismic vibration and GGN analysis}
\label{subsec:seismic-disc}

Seismic vibrations source two different systematic effects on the interferometer. Direct vibration of the laser and optics system can lead to phase noise in the same way that inherent laser noise affects the phase~\cite{abe2021matter,Asenbaum:2017,Graham:2013}. Another effect of ground motion is to generate perturbations in the initial atom cloud kinematics by vibrating the atom sources. For a gradiometer configuration of the atom interferometer the velocity difference between the two atom clouds leads to a phase shift. An example of the expected noise levels associated with initial cloud kinematics and more detail can be found in reference~\cite{abe2021matter}.

The last systematic is an effective strain that is caused by GGN. From current analysis the gradiometer phase shift measurement of the atom interferometer, assuming a GW signal, leads to a spurious strain noise of
\[
  \delta h_{\text{GGN}} = \frac{2\pi G \gamma (\nu) \rho_{0} }{L \omega^{2}_{ggn}} \ev{\delta\xi_{z}}  \left(e^{- z_{0} \frac{\omega_{ggn}}{c_{R}}} - e^{- (L+z_{0}) \frac{\omega_{ggn}}{c_R}} \right),
\]
where $z_{0}$ is the depth of the interferometer closest to the surface in the gradiometer configuration with baseline $L$. The magnitude of this strain noise is $\sim \qty(10^{-17}/\si{\sqrt{Hz}})$ for $L = \SI{100}{m}$, $z_{0} = \SI{10}{m}$, a Rayleigh wave with frequency $\omega_{ggn} = 2\pi \times\SI{1}{Hz}$ and velocity $c_{R} = \SI{210}{m/s}$, and a vertical surface displacement amplitude $\delta\xi_{z} = \SI{1}{\micro\meter}$. These Rayleigh wave values are used for all following plots. This level of strain noise will mask signals of GWs that we wish to look for in our frequency band of \SI{0.3}{Hz}--\SI{10}{Hz} with future long baseline atom interferometers. A benefit of atom interferometers is that we can use many interferometers down the baseline to distinguish GGN from GWs which cannot be done in laser based interferometers~\cite{Harms_2013}.

\begin{figure}
  \centering
  \includegraphics[height=10cm]{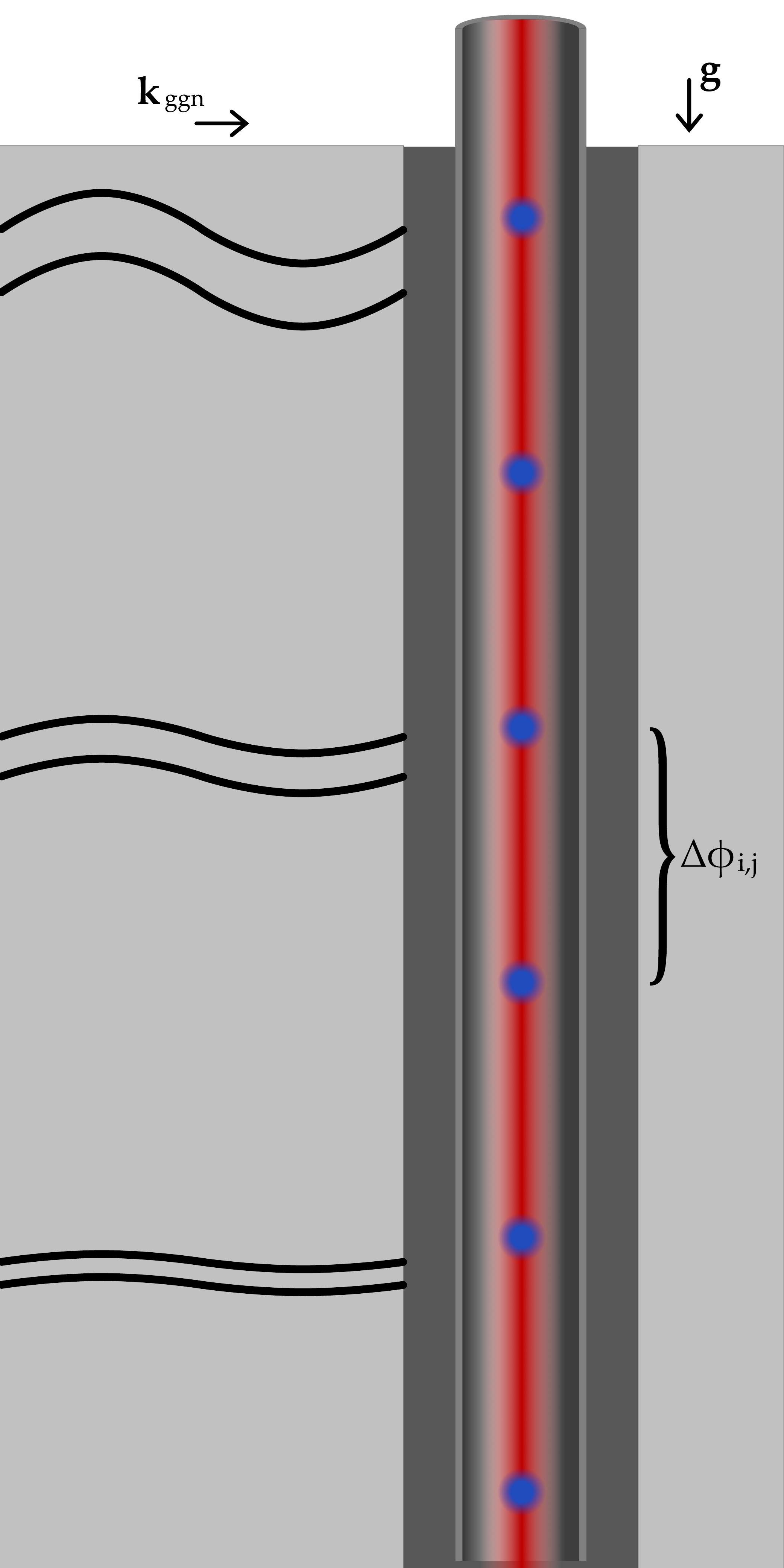}
  \caption{\label{fig:string} Diagram of vertical string-of-pearls configuration for multiple atom interferometers (blue dots) down the baseline. Gradiometer phase difference between interferometers ($\Delta\phi_{i,j}$) can be used to characterize the exponential decrease in GGN amplitude (shown as black waves).}
\end{figure}
For future suppression and mitigation of GGN we investigated the use of a string-of-pearls method, as shown in figure~\ref{fig:string}, where multiple interferometers are generated along the baseline and run simultaneously in a gradiometer configuration to probe the gravitational environment. By adding more gradiometer measurements we would expect to increase the signal to noise ratio of the GGN by increased statistics as well as gain increased information about the spatial dependence of the signal down the baseline. By exploiting the fact that GGN and GW signals are expected to have different dependencies on height, this spatially resolved detection of the signal can be used to distinguish a GGN signal from a GW signal.  Even by using three atom sources in MAGIS-100 the atom clouds created at each location can be split into multiple atom clouds using the technique demonstrated in reference~\cite{Asenbaum:2017}. A similar method is planned for MIGA and ELGAR~\cite{canuel2020,Canuel_2018}, however in our scenario the baseline is vertical and the atom trajectories are perpendicular to the surface of the Earth, whereas MIGA and ELGAR use a horizontal vacuum pipe with the atom clouds spread out horizontally along the baseline and launched vertically. Each atom interferometer in their configuration is thus at the same distance away from the surface where the sourcing Rayleigh waves occur. We then measure the phase shift for a \SI{10}{m} launch at different heights along the baseline. In this simulation we generated 5 atom clouds per atom source, 15 atom clouds in total for a \SI{100}{m} baseline (MAGIS-100) and 20 atom clouds for a \SI{1}{km} baseline assuming 1000 $\hbar k$ LMT atom optics. With some technical advancement this scheme can be realized in the MAGIS-100 detector as a proof of concept. A plane Rayleigh surface wave was assumed with velocity of \SI{210}{m/s} and a vertical displacement amplitude of \SI{e-6}{m} in agreement with seismometer measurements around the shaft. The phase shift of the detector was calculated using a perturbative semi-classical method~\cite{hogan2008}. The resulting GGN phase shift in the atom interferometer to leading order has the form
\begin{equation}
  \label{eq:5}
  \delta\phi_{\text{GGN}} = \frac{8\pi G \gamma (\nu) \rho_{0} }{\omega^{2}_{ggn}} \ev{\delta\xi_{z}} nk_{eff} e^{-k_{l} z} \sin^{2}\left(\frac{\omega_{ggn} T}{2}\right)\cos(\phi_{ggn}-\omega_{ggn} T).
\end{equation}
which is exponential with depth $z$, while the response of the interferometer to GWs is linear with the baseline $L$~\cite{DIMOPOULOS2009}
\begin{equation}
  \label{eq:7}
  \delta\phi_{\text{GW}} = 4k_{eff}hL\sin^{2}\qty(\frac{\omega_{gw}T}{2}) \sin(\phi_{0}).
\end{equation}

\begin{figure}
  \centering
  \includegraphics[width=0.6\textwidth]{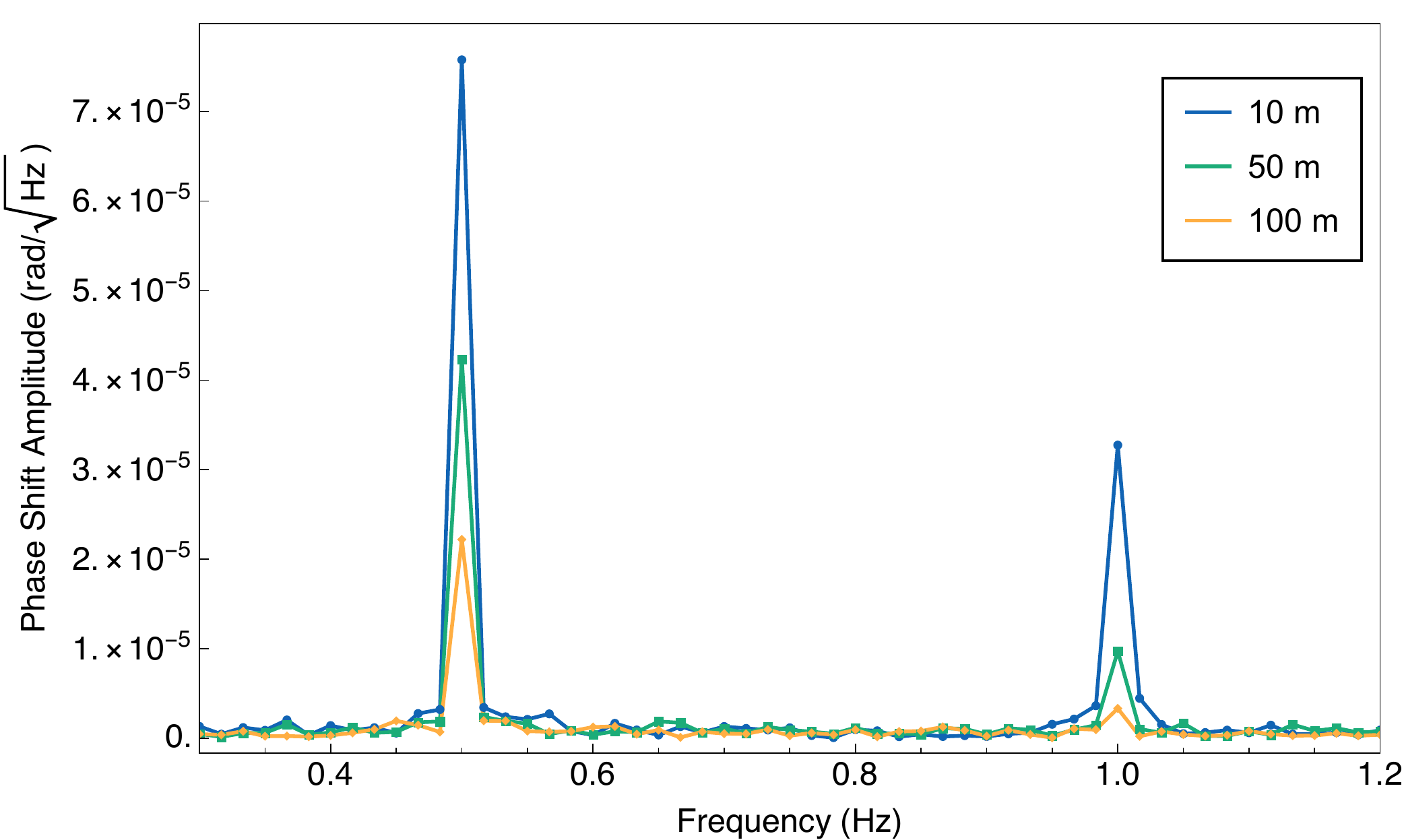}
  \caption{\label{fig:2freqFFT} Example FFT of sampled GGN phase shift at 10~m, 50~m, and 100~m along the baseline. Data sampled with a rate of \SI{10}{Hz} and sample period of \SI{60}{s} and 10\% noise.}
\end{figure}

\begin{figure}
  \centering
  \includegraphics[width=\textwidth]{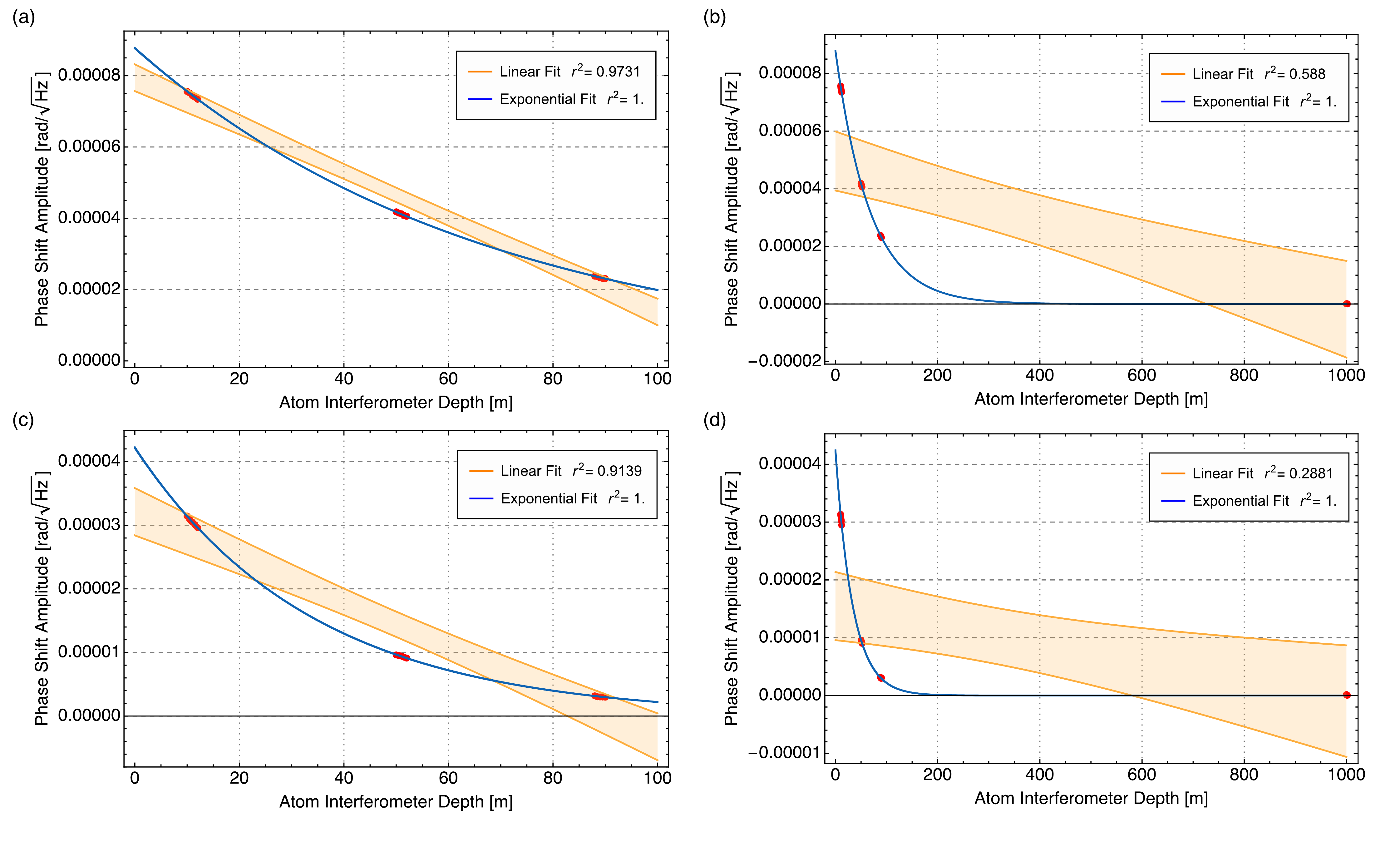}
  \caption{\label{fig:fft-fitgrid} Fits of GGN phase shift Fourier amplitude at various depths along the atom interferometer baseline with phase noise of 1\% of the maximum GGN signal added. (a, b) are fits of \SI{0.5}{Hz} peak at \SI{100}{m} and \SI{1}{km} baselines respectively. (c, d) are fits of \SI{1}{Hz} peak. }
\end{figure}

To analyze the ability to fit the GGN exponential form, a simulation of the Fourier analysis phase extraction was done. A time dependent GGN phase signal, $\delta\phi_{\text{GGN}} (t)$ where $\phi_{ggn} = (\psi_{ggn} - \omega_{ggn}t)$ from equation~\ref{eq:5}, was generated with two Rayleigh wave frequencies and phase noise to mimic a sequence of atom interferometer measurements for specific sampling rates and periods at different heights down the baseline. This simulated data was then transformed with a Fast-Fourier-Transform (FFT)~\cite{brigham1988fast}, see figure~\ref{fig:2freqFFT}. The frequency peaks were found and the corresponding peak amplitudes were extracted for fitting to a linear model and an exponential model, as shown in figure~\ref{fig:fft-fitgrid}.

To understand the effectiveness of this method on reducing GGN noise from a gradiometer phase difference measurement, we define a suppression factor as the ratio of the linear phase difference across the baseline\footnote{ Calculated from the linear term of a mixed model fit of the form $A e^{-\alpha z} + B z$, where $A$ is the amplitude fit parameter, $\alpha$ is the exponential fit parameter, and $B$ is the linear fit parameter} and a simulated phase difference measurement, 
\begin{equation}
\label{eq:8}
\text{suppression factor} = \frac{\sqrt{\Delta\phi_{linear}(z_1,z_2)^2 + \sigma_{linear}^2}}{\Delta\phi_{sim}(z_1,z_2)},
\end{equation}
where $\Delta\phi_{sim}(z_1,z_2) \equiv \abs{\phi_{sim}(z_{1}) - \phi_{sim}(z_{2})}$ is the simulated phase difference with phase noise between interferometers at $z_1$ and $z_2$, $\Delta\phi_{linear}(z_1,z_2)$ is the phase difference between $z_1$ and $z_2$ calculated from the linear term of the mixed model fit, and $\sigma_{linear}$ is the standard error for the linear slope multiplied by the difference of $z_{1}$ and $z_{2}$. The result of this is a residual linear term which acts as noise in GW and DM measurements. The suppression factor corresponds to the ratio of the residual GGN-induced linear signal that arises in the mixed model fit to the linear GGN signal that would be naively inferred by performing a linear extrapolation between the two points on either end of the baseline. In this study we assumed a phase noise at various percentages of the maximum GGN signal amplitude. This could arise from shot noise or other technical sources~\cite{abe2021matter}.

\begin{figure}
  \centering
  \includegraphics[width=\textwidth]{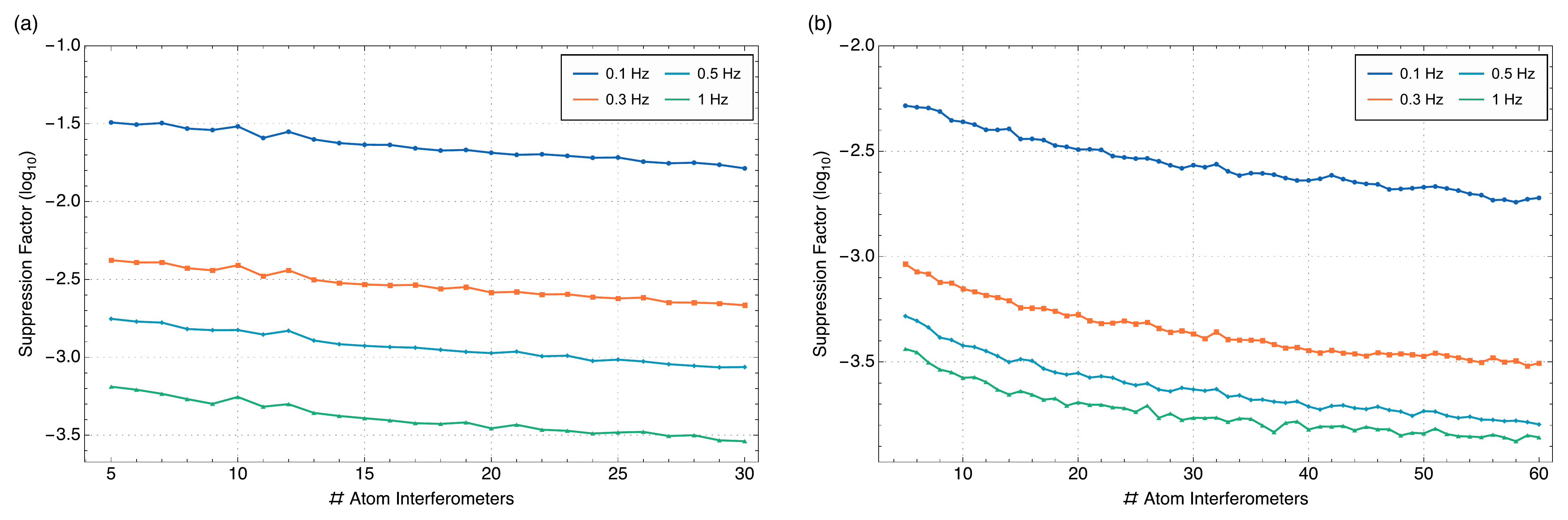}
  \caption{Suppression factor of GGN noise versus number of atom interferometers down the baseline for (a) \SI{100}{m} and (b) \SI{250}{m}. Plots show the effect for a Rayleigh wave with a frequency: \SI{0.1}{Hz}, \SI{0.3}{Hz}, \SI{0.5}{Hz}, and \SI{1}{Hz}. A sample period of \SI{60}{s} and sample rate of \SI{10}{Hz} is used for the signal. Suppression factor amplitudes are RMS values calculated after 1000 sets of fits each with randomized Gaussian phase noise of 1\% of the maximum GGN signal amplitude at the surface. \label{fig:suppvsnumAI}}
\end{figure}

\begin{figure}
  \centering
  \includegraphics[width=0.6\textwidth]{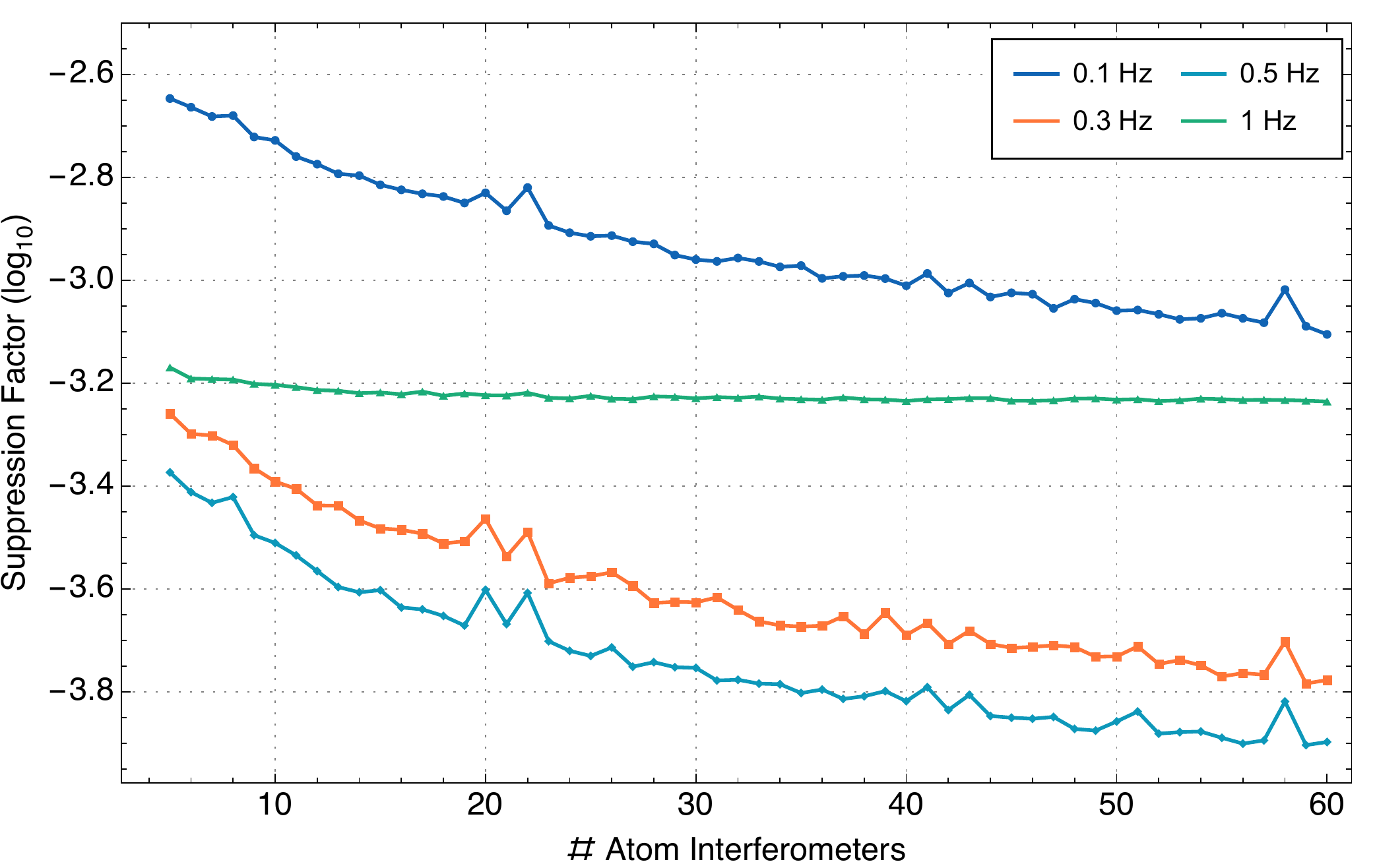}
  \caption{\label{fig:supp-400m} Suppression factor of GGN noise versus number of atom interferometers down the baseline for \SI{400}{m}. Plot shows the effect for a Rayleigh wave with a frequency: \SI{0.1}{Hz}, \SI{0.3}{Hz}, \SI{0.5}{Hz}, and \SI{1}{Hz}. A sample period of \SI{60}{s} and sample rate of \SI{10}{Hz} is used for the signal. Suppression factor amplitudes are RMS values calculated after 1000 sets of fits each with randomized Gaussian phase noise of 1\% of the maximum GGN signal amplitude at the surface.}
\end{figure}

\begin{figure}
  \centering
  \includegraphics[width=\textwidth]{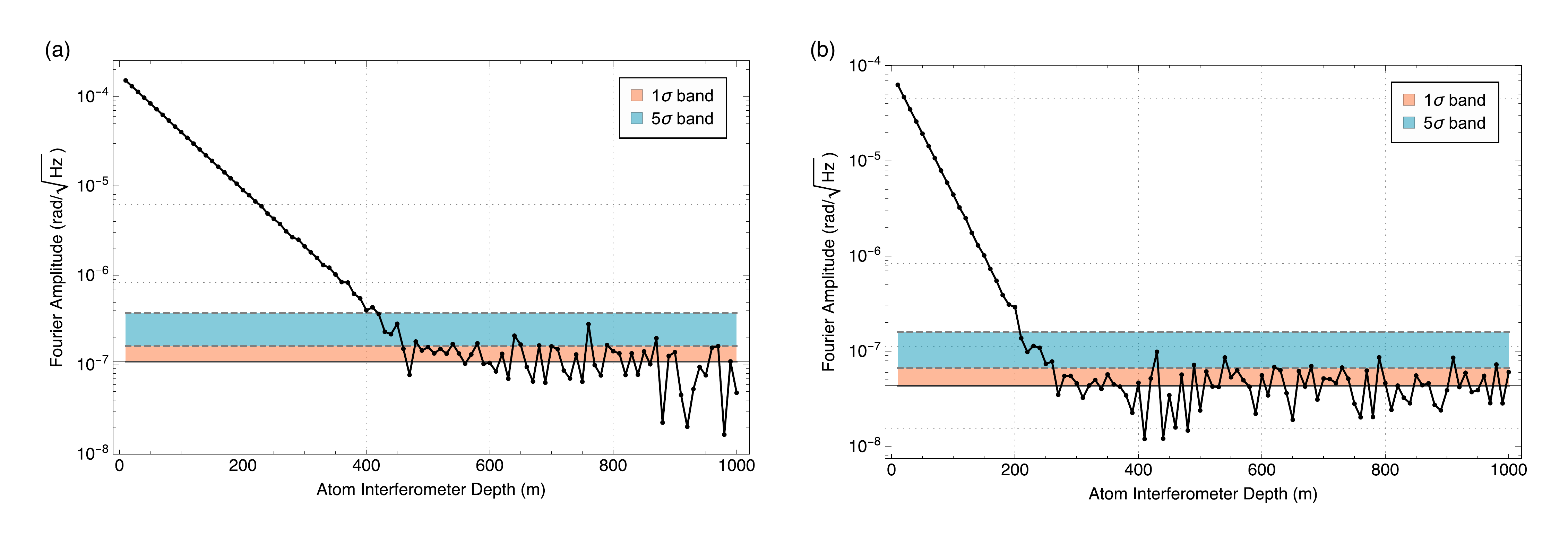}
  \caption{\label{fig:amplitudenoisefloor} Fourier amplitudes extracted for Rayleigh waves of (a) \SI{0.5}{Hz} and (b) \SI{1}{Hz}. 1$\sigma$ and 5$\sigma$ bands are number of standard deviations from the mean of the noise floor in the FFT.}
\end{figure}

The suppression factor was compared with increasing the measurement resolution along the baseline by using multiple, equally spaced, interferometers for each fit, as shown in figure~\ref{fig:suppvsnumAI}. The figure shows the suppression factor decreasing approximately as $1/\sqrt{n}$ where $n$ is the number of interferometers along the baseline. With no phase noise present we found a perfect suppression on the order of $10^{-15}$, which is the machine precision of the computer the simulation was run on. Figure \ref{fig:amplitudenoisefloor} shows the extracted Fourier amplitudes as a function of depth and also shows standard deviations $\sigma$ from the mean of the noise floor in the FFT. The GGN signal falls into the 1$\sigma$ band below \SIrange{250}{400}{m}, depending on the Rayleigh wave frequency. Increasing the number of interferometers below this depth would not help in pushing down the residual linear term of the mixed model fit since those amplitudes are noise limited. This is clearly evident for the Rayleigh wave with frequency of \SI{1}{Hz} in figure~\ref{fig:supp-400m} where the suppression factor is calculated using data points in the 1$\sigma$ band of the noise floor (figure~\ref{fig:amplitudenoisefloor}b). So, for GGN calibration and \textit{in-situ} measurement data above approximately \SI{250}{m} will be most useful in mapping out the exponential character for all frequencies, while the entire baseline will be used for GW and DM measurements as these signals are linear and maximized with longer baseline.

\begin{figure}
  \centering
  \includegraphics[width=\textwidth]{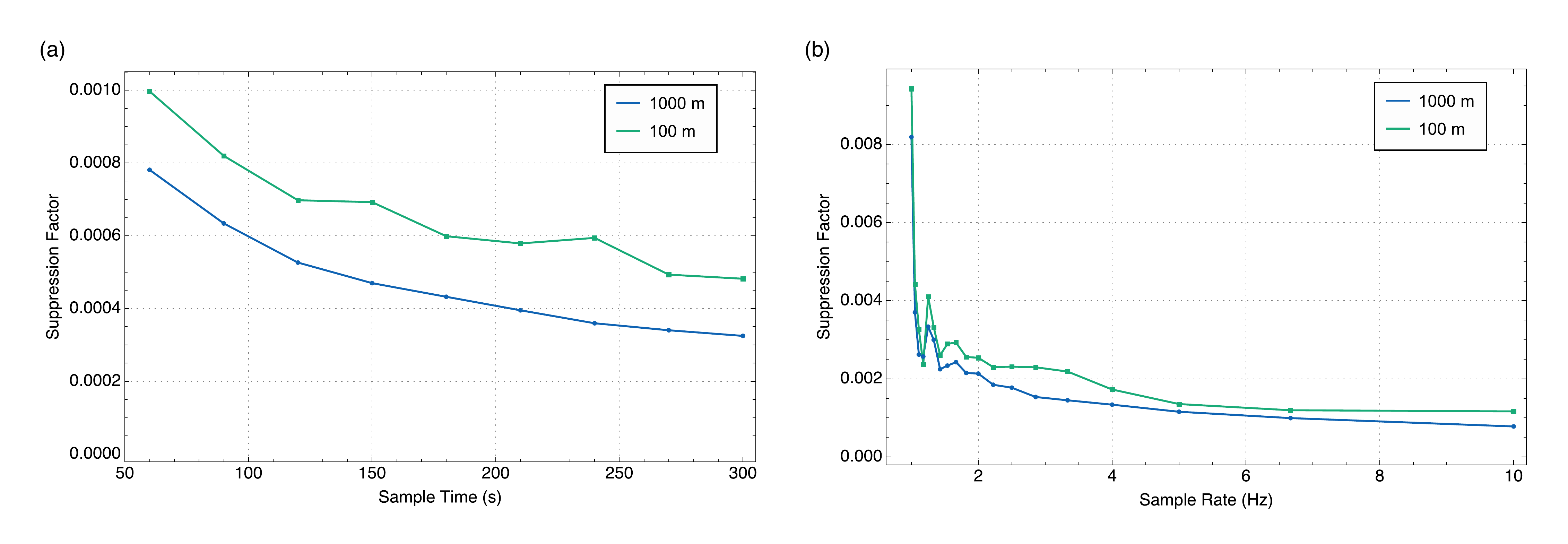}
  \caption{\label{fig:supp-params} Dependence of the effects of varying sample period and sample rate on a suppression factor for 20 equally spaced atom interferometers down the baselines of 100~m and 1000~m. Suppression factor values are RMS calculated for 100 sets of fits with randomized Gaussian phase noise. (a) Plotted with phase noise of 1\% and sample rate of \SI{10}{Hz}. (b) Plotted with phase noise of 1\% and a sample period of \SI{60}{s}}
\end{figure}

\begin{figure}
  \centering
  \includegraphics[width=0.6\textwidth]{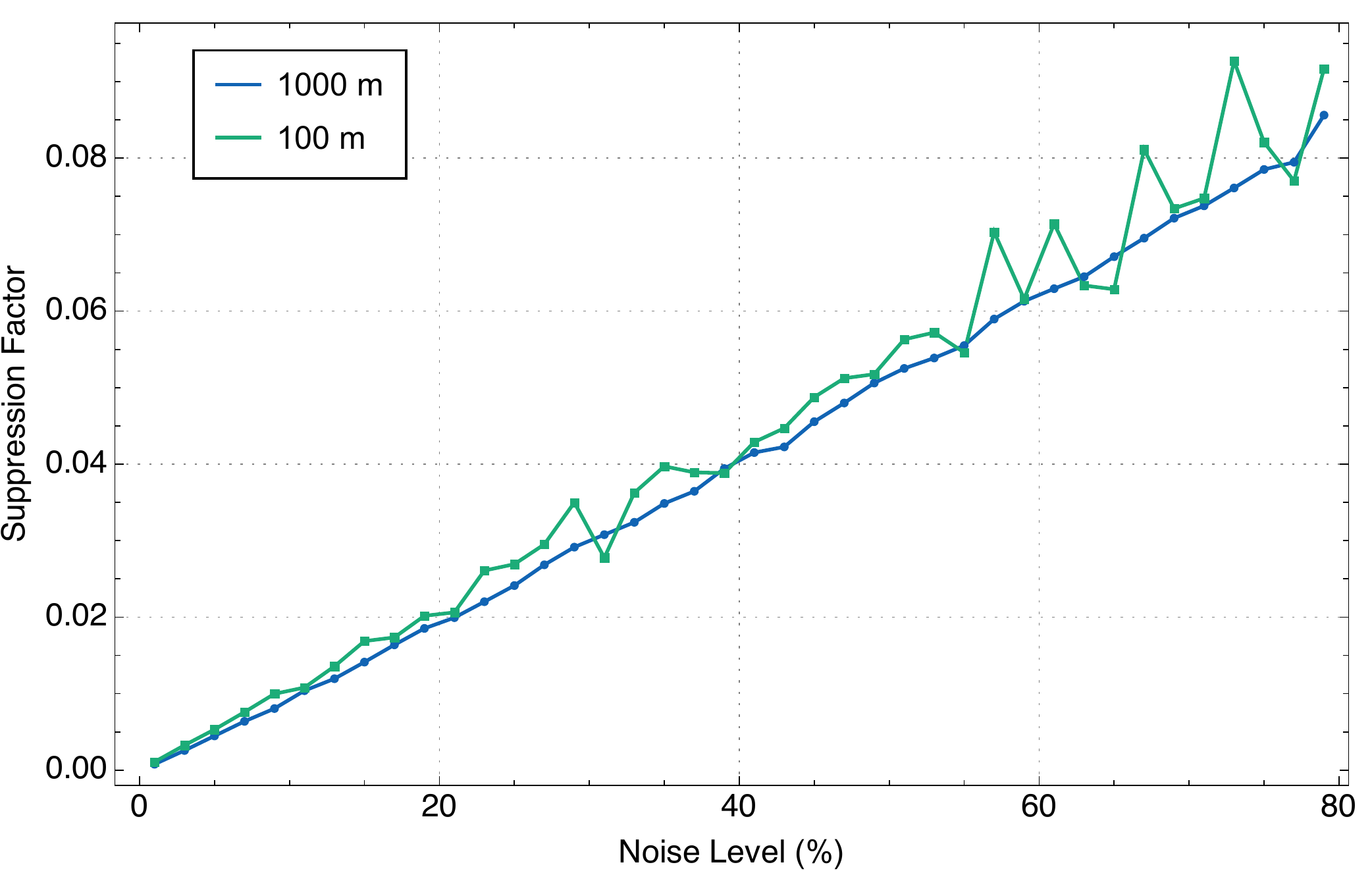}
  \caption{\label{fig:supp-noiselvl} Effect of varying noise level as a percentage of the maximum GGN backround amplitude on suppression factor for 20 equally spaced atom interferometers down the baseline. Suppression factor values shown are RMS of 100 sets of fits with randomized Gaussian phase noise, sample period of \SI{60}{s}, sample rate \SI{10}{Hz}.}
\end{figure}

Effects of varying the simulation parameters on the suppression factor were then explored for fits of 20 atom interferometers equally spaced along the baseline, see~\cref{fig:supp-params,fig:supp-noiselvl}. These plots show that each of the parameters: phase noise level, sample period, and sample rate all have a direct impact on the suppression factor with the phase noise level having the largest effect, however, even in the worst case scenarios there is significant suppression achieved.

\begin{figure}
  \centering
  \includegraphics[width=0.6\textwidth]{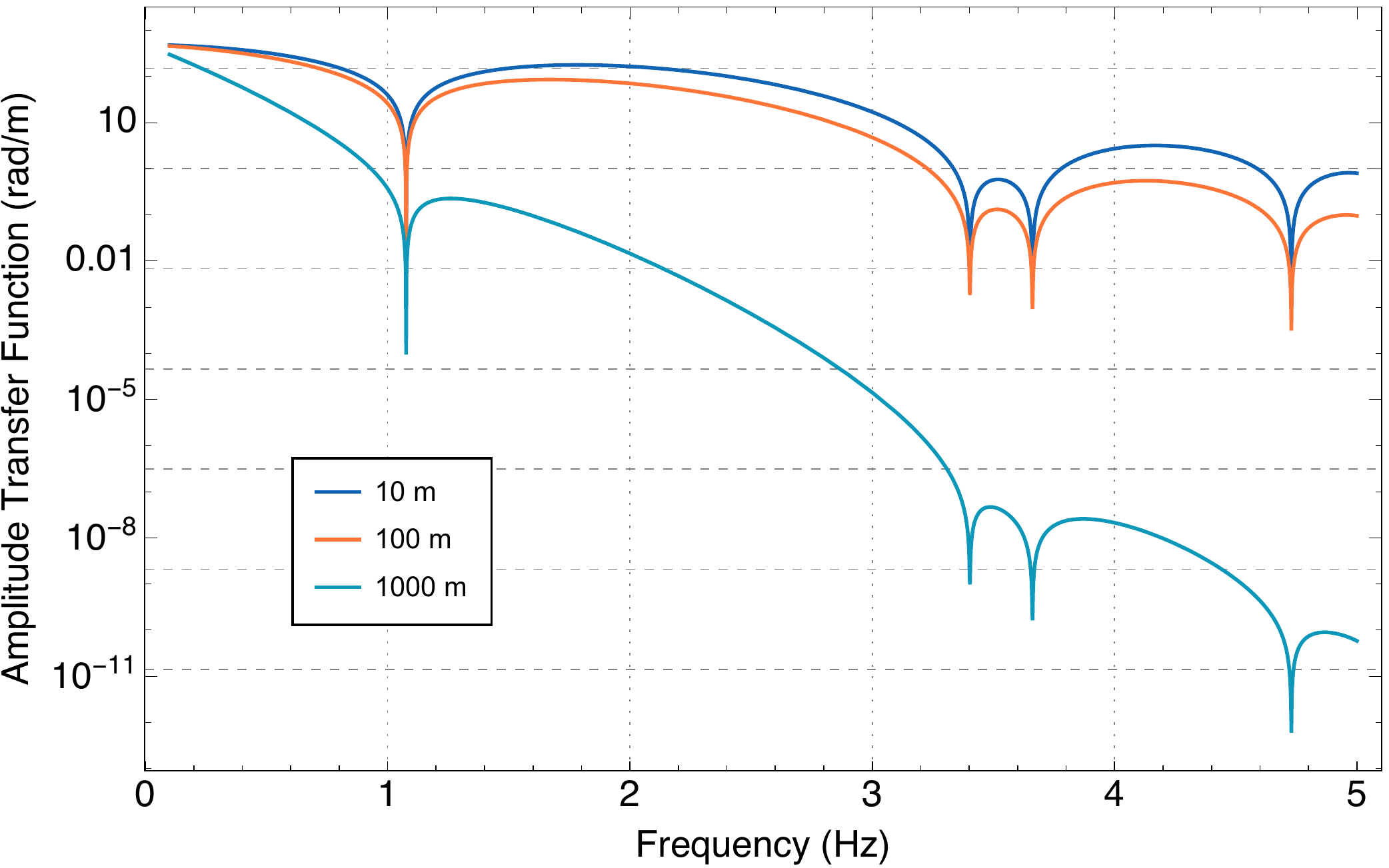}
  \caption{Analytical transfer functions of the atom interferometer phase response to a vertical ground displacement plotted over Rayleigh wave frequencies. Three atom interferometer's distances from the surface are shown.}
  \label{fig:ggntransfer}
\end{figure}

Finally, we examined the amplitude transfer function $\dv*{\phi_{\mathrm{GGN}}}{(\delta\xi_{z})}$ between a single atom interferometer phase response $\phi_{\mathrm{GGN}}$\footnote{$\phi_{\mathrm{GGN}}$ is the full phase shift of the interferometer including the leading order term in eq.~(\ref{eq:5}) and higher orders.} and vertical ground displacement $\delta \xi_{z}$ at the surface of the shaft, shown in \cref{fig:ggntransfer}. We plotted the transfer function for various locations of the atom interferometer underground and over the Rayleigh wave frequencies using the same experimental parameters as above. The farther the atom interferometer is from the surface the smaller the impact of the GGN phase shift, especially from higher frequency Rayleigh wave sources. Combining active seismometer arrays~\cite{Harms:2015}, to provide priors, with \textit{in-situ} calibration measurements using the atom interferometer will allow us to map out the GGN and suppress it from the atom interferometer signal. Further studies will be needed to investigate the effect of other seismic waves and ground inhomogeneities on the atom interferometer response.

The secondary seismic vibration effect of GGN as shown in~\cref{fig:ggn-model} is not an early systematic for the MAGIS-100 detector as we will not reach the strain sensitivity to detect it until technical advancements have been made. Active monitoring will be used for seismic vibration mitigation as well as early diagnostic measurements. Once the MAGIS-100 detector has reached an advanced stage we may be sensitive enough to reach strains where GGN becomes detectable. We would be interested to both study as well as suppress the effect of GGN on our measurements as it interferes with GW and DM detection. In addition to GGN generated by surface seismic waves the atmospheric pressure fluctuations also act as a source for gravity gradients detectable by atom interferometers~\cite{Creighton_2008}. Further analysis will collect data on the pressure fluctuations near the site to determine what level this systematic will have. 

\subsubsection{Future outlook}\label{sec:outlook}

Future studies of GGN in longer baselines will consider the effect of higher velocity Rayleigh waves deeper underground, where the dense rock supports faster traveling waves. Underground there will also be contributions from body waves, specifically P- and S-waves, that can become surface waves between differing mediums after scattering that will also contribute to the seismic perturbation spectrum. In addition the longer baseline will contribute directly to reducing the impact of the GGN on strain measurements.

\section{Conclusions}
\label{sec:conclusions}

MAGIS-100 aims to be an extremely sensitive large baseline atom interferometer pathfinder. It will search for ultra-light DM and serve as a prototype for exploring the mid-band frequency spectrum for GW signals. For any terrestrial atom interferometer with these goals, GGN is a critical background that limits access to lower frequencies. Since the GW and DM detection channels have signals that vary linearly with height along the vertical baseline, whereas the GGN background from Rayleigh waves exponentially decays with depth, correlations between multiple atom interferometers at different heights can be leveraged to perform \textit{in-situ} characterization of GGN and to correspondingly suppress the extent to which the GGN background leaks into the GW and DM detection channels by up to orders of magnitude. Our simulation motivates further studies into mitigation strategies using correlations between multiple interferometers along the baseline as well as correlations with seismometer arrays at the surface. With a more robust understanding of the effects of GGN on large baseline atom interferometers we hope for the ability to access even lower frequency ranges to expand our science reach.

We have characterized the temperature and seismic environment of the MAGIS-100 installation site and examined the expected magnitudes of the systematic noise that arises from environmental fluctuations. We plan to actively monitor the environment around the shaft to further study the impact and mitigation strategies being developed for MAGIS-100 noise sources. These studies have and will continue to guide research and development of various crucial engineering and design choices for the optical system, the atom sources, and the vacuum pipe.

Further investigations include a complete year of temperature, humidity, and seismic monitoring as well as measurement of barometric pressure fluctuations in order to infer atmospheric Newtonian noise~\cite{Saulson:1984}. Studies of alternative mitigation strategies for GGN at a baseline of \SI{1}{km} are also ongoing. Future analysis of GGN at extreme baseline lengths will include Rayleigh wave velocity dependence on depth as well as other contributing terms from seismic body waves underground. Lower frequency GGN should be measurable by MAGIS-100 earlier on and be useful for geophysics research. There is also great interest in exploring geophysical processes with atom interferometers~\cite{Canuel_2018}.

\section{Acknowledgments}
\label{sec:acknowledgments}

We would like to thank our collaborators Jason Hogan, Peter Graham, Roni Harnik, and Jon Coleman for insightful discussions on the technical and theoretical sides of this analysis. Also we are grateful to Philippe Bouyer and Ernst Rasel for their conversations and input. This project is funded in part by the Gordon and Betty Moore Foundation Grant GBMF7945. This document was prepared using the resources of the Fermi National Accelerator Laboratory (Fermilab), a U.S. Department of Energy, Office of Science, HEP User Facility. Fermilab is managed by Fermi Research Alliance, LLC (FRA), acting under Contract No. DE-AC02-07CH11359. This work is supported in part by the U.S. Department of Energy, Office of Science, QuantiSED Intitiative. Jeremiah Mitchell acknowledges support from the Kavli Foundation as well as the Presidential research grant from Northern Illinois University.

\bibliography{bibliography}

\end{document}